%% file: HCVS25-ProofDeductionSatisfaction-CameraReady-sent.tex
\documentclass[preliminary,copyright,creativecommons]{eptcs}
\providecommand{\event}{HCVS 2025} 

\usepackage{hyperref}

\input{packages}

\input{macros}

\title{Semantic Properties of Computations Defined by Elementary Inference Systems%
\thanks{Supported by 
project
PID2021-122830OB-C42 funded by
MCIN/AEI/10.13039/501100011033 and by ``ERDF A way of making Europe''
and by the grant CIPROM/2022/6 funded by Generalitat Valenciana
} 
}
\author{Salvador Lucas 
\institute{DSIC \& VRAIN, Universitat Polit\`ecnica de Val\`encia, Spain}
\email{slucas@dsic.upv.es}
}

\newcommand{\titlerunning}{Semantic Properties of Computations Defined by Elementary Inference Systems}
\newcommand{\authorrunning}{Salvador Lucas}

\hypersetup{
  bookmarksnumbered,
  pdftitle    = {\titlerunning},
  pdfauthor   = {\authorrunning},
  pdfsubject  = {EPTCS},               
  pdfkeywords = {First-Order Logic, Program Analysis, Rewriting Systems} 
}

\begin{document}
\maketitle

\begin{abstract}
We consider sets/relations/computations defined by \emph{Elementary Inference Systems} $\GLinference$, which are obtained from Smullyan's \emph{elementary formal systems} using Gentzen's notation for inference rules, and proof trees for 
atoms $P(t_1,\ldots,t_n)$, where predicate $P$ represents the considered set/relation/computation.
A first-order theory $\GLtheoryOf{\GLinference}$, actually a set of definite Horn clauses, is given to $\GLinference$.
Properties of objects defined by $\GLinference$ are expressed as first-order sentences $\GLformula$, 
which are proved true or false by \emph{satisfaction} 
$\satisfactionInThOf{\StdModel{}}{\GLformula}$ of $\GLformula$ in a \emph{canonical} model 
$\StdModel{}$
of $\GLtheoryOf{\GLinference}$.
For this reason, we call $\GLformula$ a \emph{semantic property} of $\GLinference$.
Since canonical models are, in general, incomputable, we show how to (dis)prove semantic properties  by satisfiability in an \emph{arbitrary} model $\SStructure$ of $\GLtheoryOf{\GLinference}$.
We apply these ideas to the analysis of properties of programming 
languages and systems whose computations can be described by means of an elementary inference system. In particular, rewriting-based systems.
\end{abstract}

\section{Introduction}\label{SecIntroduction}

Elementary formal systems \cite{Smullyan_TheoryOfFormalSystems_1961} provide an appropriate
 device for the definition and combination 
 of sets, relations, and hence of computational relations, which is
 amenable for \emph{mechanization}.
The \emph{operational semantics} of computational systems and programming languages is 
often given by 
means of a formal system, usually presented as a set of \emph{inference rules} which are used to 
\emph{prove}
goals $P(t_1,\ldots,t_n)$ for some predicate symbol $P$ 
(representing the considered set of elements or tuples of elements) 
and terms $t_1,\ldots,t_n$ 
(representing components or tuples of components).
In \cite{Plotkin_TheOriginsOfStructuralOperationalSemantics_JLAP04}, 
Plotkin recalls the role of Smullyan's formal systems \cite{Smullyan_TheoryOfFormalSystems_1961} in the development of his Structural Operational Semantics (\sos{}
\cite{Plotkin_AStructuralApproachToOperationalSemantics_DAIMI81, Plotkin_AStructuralApproachToOperationalSemantics_JLAP04}),\footnote{However, \cite{Plotkin_AStructuralApproachToOperationalSemantics_JLAP04} contains no reference to Smullyan.}
 which is widely used in the semantic description of programming languages since the 1980s, see, e.g., 
 \cite{Kahn_NaturalSemantics_STACS87}.
Plotkin also mentions Barendregt's PhD thesis \cite{Barendregt_SomeExtensionalTermModelsForCombinatoryLogicsAndLambdaCalculi_PhD71} where $\lambda$-calculus is described using inference rules, see \cite[Appendix I]{Barendregt_SomeExtensionalTermModelsForCombinatoryLogicsAndLambdaCalculi_PhD71}. In particular, he displays this rule (from \cite[page 12]{Barendregt_SomeExtensionalTermModelsForCombinatoryLogicsAndLambdaCalculi_PhD71}):
\begin{eqnarray}
\bigfrac{N\triangleright N'}{M~N\triangleright M~N'}\label{LblPropagationBetaReduction2}
\end{eqnarray}
where, as in \cite{HinSel_IntroductionToCombinatorsAndLambdaCalculus_1986}, we use $\triangleright$ instead of Barendregt's original $\geq$ to denote $\beta$-reduction.
Rule (\ref{LblPropagationBetaReduction2}) expresses that $\beta$-reduction is \emph{propagated} on the \emph{second} argument of $\lambda$-calculus \emph{application} (with binary operator $\_~\_$). There is a similar rule for propagation on the first argument as well.

Despeyroux introduced the term \emph{Natural Semantics} \cite{Despeyroux_ProofOfTranslationInNaturalSemantics_LICS86} to refer to \emph{the purely `formal system' part} of \sos{} which actually relies on Gentzen's Natural Deduction \cite{Gentzen_UntersuchungenUberDasLogischeSchliessenPart1_MZ35,Prawitz_NaturalDeductionAProofTheoreticalStudy_1965}, where
proofs of computations are represented by means of \emph{proof trees}.
In order to \emph{reason} about computations described with such formal systems, the use of
\emph{first-order formulas} which can be proved true or false of the defined object is a natural choice
to \emph{express} properties \cite[Section 1.1, last paragraph]{Kahn_NaturalSemantics_STACS87}.
As posed by Kahn,
\begin{quote}
A semantic definition is a list of axioms and inference rules that define predicates.
A semantic definition is identified with a logic, and reasoning with the language is proving theorems within that logic \cite[page 23, third paragraph]{Kahn_NaturalSemantics_STACS87}.
\end{quote}
We essentially subscribe this point of view, although the ``reasoning as theorem proving'' part will be revisited.

For instance, the operational description of one-step 
reduction
$\rew{\cR}$ 
 in reduction-based systems $\cR$ 
 allowing for \emph{conditional rules} $\lhsr\to\rhsr\IF\gencond$
 is naturally made by using inference rules \cite{BruMes_SemanticFoundationsForGeneralizedRewriteTheories_TCS06,%
Lalement_ComputationAsLogic_1993,%
Lucas_AnalysisOfRewritingBasedSystemsAsFirstOrderTheories_LOPSTR17,%
Meseguer_ConditionalRewritingLogicAsAUnifiedModelOfConcurrency_TCS92,%
Meseguer_TwentyYearsOfRewritingLogic_JLAP12}.
 One-step rewriting is defined as \emph{provability} 
 of goals $s\rew{}t$, 
where (as in \cite{Despeyroux_ProofOfTranslationInNaturalSemantics_LICS86})
the usual rewriting symbol $\rew{}$ is viewed as a predicate symbol,
in an inference system $\GLinferenceOf{\cR}$.
 We illustrate this with \emph{Generalized Term Rewriting Systems} (\gtrs{s} 
\cite{Lucas_LocalConfluenceOfConditionalAndGeneralizedTermRewritingSystems_JLAMP24}) which generalize Conditional Term Rewriting Systems (\ctrs{s} \cite{Kaplan_ConditionalRewriteRules_TCS84}) by enabling the use of
  \emph{atoms} in the conditions of rules, possibly defined by definite Horn clauses which are part of the \gtrs.
It is also possible to establish which arguments 
of each $k$-ary function symbol $f$ can be rewritten
by means of a \emph{replacement map} $\mu$ which specifies them as a set $\mu(f)\subseteq\{1,\ldots,k\}$ of \emph{active} arguments \cite{Lucas_ContextSensitiveRewriting_CSUR20}.
In particular, $\muBot$ \emph{forbids} reductions in all arguments of all function symbols, i.e., $\muBot(f)=\emptyset$.
\begin{example}\label{ExPEvenZeroOdd}
The \gtrs{} $\cR=(\Symbols,\SPredicates,\mu,H,R)$, 
with 
$\Symbols=\{\fS{0},\fS{s}\}$, 
$\SPredicates=\{\rew{}, \rews{}, \geq, \fS{odd},  \fS{peven},\fS{zero}\}$,
$\mu=\mu_\bot$,
$H=\{(\ref{ExPEvenZeroOdd_horn1}),(\ref{ExPEvenZeroOdd_horn2}),(\ref{ExPEvenZeroOdd_horn3}),(\ref{ExPEvenZeroOdd_horn4}),(\ref{ExPEvenZeroOdd_horn5})\}$,
and $R=\{(\ref{ExPEvenZeroOdd_rule1})\}$, where:\\
\noindent
\begin{tabular}{c@{\hspace{0.8cm}}c}
\begin{minipage}{7.5cm}
\begin{eqnarray}
x  \geq  0&& \label{ExPEvenZeroOdd_horn1}\\
\fS{s}(x) \geq \fS{s}(y) & \IF & x\geq y\label{ExPEvenZeroOdd_horn2}\\
\fS{peven}(x) & \IF & x\rews{}\fS{s}(\fS{s}(\fS{0}))\label{ExPEvenZeroOdd_horn3}
\end{eqnarray}
\end{minipage}
&
\begin{minipage}{7.5cm}
\begin{eqnarray}
\fS{odd}(x) & \IF & x\rews{}\fS{s}(\fS{0})\label{ExPEvenZeroOdd_horn4}\\
\fS{zero}(x) & \IF & x\rews{}\fS{0}\label{ExPEvenZeroOdd_horn5}\\
\fS{s}(\fS{s}(x)) \to   x &  \IF & x\geq\fS{s}(\fS{0})\label{ExPEvenZeroOdd_rule1}
\end{eqnarray}
\end{minipage}
\end{tabular}

\bigskip
\noindent
can be used to \emph{classify} natural numbers $n\in\naturals$ written in Peano's notations, i.e., as $\fS{s}^n(\fS{0})$, into \emph{odd}, \emph{positive and even}, or 
\emph{zero} by using predicate symbols $\fS{odd}$, $\fS{peven}$, and $\fS{zero}$, respectively.
Predicate 
$\geq$ is defined by the Horn clauses (\ref{ExPEvenZeroOdd_horn1}) and (\ref{ExPEvenZeroOdd_horn2});
clauses (\ref{ExPEvenZeroOdd_horn3}), (\ref{ExPEvenZeroOdd_horn4}), and (\ref{ExPEvenZeroOdd_horn5}) define the tests;
and rule (\ref{ExPEvenZeroOdd_rule1}) defines one-step rewriting.
Computations with $\cR$ can be defined by the elementary inference system
$\GLinferenceOf{\cR}$  in 
Figure \ref{FigExPEvenZeroOdd_IS}.
\begin{figure}[t]
\begin{tabular}{cc@{\qquad\qquad\quad}cc@{\qquad\qquad\quad}cc}
$(\RuleReflexivity)$
& 
$\bigfracLabel{}{x \rightarrow^{\ast} x}$
&
$(\RuleCompatibility)$ 
& 
$\bigfracLabel{x \rightarrow y\qquad y
  \rightarrow^{*} z}{x  
\rightarrow^{*} z}$
&
$(\RuleHornClause)_{(\ref{ExPEvenZeroOdd_horn1})}$
& 
$\bigfracLabel{}{x  \geq  0}$ 
\\[0.5cm]
$(\RuleHornClause)_{(\ref{ExPEvenZeroOdd_horn2})}$
& 
$\bigfracLabel{x\geq y}{\fS{s}(x) \geq \fS{s}(y)}$
&
$(\RuleHornClause)_{(\ref{ExPEvenZeroOdd_horn3})}$
& 
$\bigfracLabel{x\rews{}\fS{s}(\fS{s}(\fS{0}))}
{\fS{peven}(x)}$
&
$(\RuleHornClause)_{(\ref{ExPEvenZeroOdd_horn4})}$
& 
$\bigfracLabel{x\rews{}\fS{s}(\fS{0})}
{\fS{odd}(x)}$
\\[0.5cm]
$(\RuleHornClause)_{(\ref{ExPEvenZeroOdd_horn5})}$
& 
$\bigfracLabel{x\rews{}\fS{0}}
{\fS{zero}(x)}$ 
&
$(\RuleHornClause)_{(\ref{ExPEvenZeroOdd_rule1})}$
& $\bigfracLabel{ x\geq\fS{s}(\fS{0})}
{\fS{s}(\fS{s}(x)) \to   x}$ 
\end{tabular}
\caption{Elementary inference system $\GLinferenceOf{\cR}$ for $\cR$ in Example \ref{ExPEvenZeroOdd}}
\label{FigExPEvenZeroOdd_IS}
\end{figure}
\end{example}
Computational properties of such 
systems $\cR$ 
are often formulated as \emph{questions} 
about the relationship between \emph{subject} expressions (e.g., terms 
$s,t,\ldots$) and the reduction relation $\rew{\cR}$ (or some of its extensions and/or combinations: $\rews{\cR}$, $\rewp{\cR}$, etc.).
Expressing such properties as first-order logic formulas is a natural choice.
A careful consideration reveals some difficulties, though.

\begin{example}
\label{ExPEvenZeroOdd_DifferencesInCanonicalModels}
For $\cR$ in Example \ref{ExPEvenZeroOdd} and $\GLinferenceOf{\cR}$ in Figure 
\ref{FigExPEvenZeroOdd_IS} the following sentence intuitively 
asserts that every number encoded as a term $\fS{s}^n(\fS{0})$ for some $n\geq 0$ 
is odd, or positive and even, or zero:
\begin{eqnarray}
(\forall x)\:\fS{odd}(x)\vee\fS{peven}(x)\vee\fS{zero}(x)\label{ExPEvenZeroOdd_TermsEitherPositiveOddZero}
\end{eqnarray}
Note that this is true \emph{only if $x$ ranges over ground terms} $t$ as above.
For instance, if $t$ is a variable $x$, then there is no proof tree
in $\GLinferenceOf{\cR}$
neither for 
$\fS{odd}(t)$, nor
$\fS{peven}(t)$, nor
$\fS{zero}(t)$, i.e., (\ref{ExPEvenZeroOdd_TermsEitherPositiveOddZero}) does \emph{not} hold.
\end{example}

Following Clark \cite{Clark_PredicateLogicAsAComputationalFormalism_TR79},
and different from Kahn (see above), 
properties of computational systems (e.g., $\cR$)
expressed as first-order sentences $\GLformula$ should be referred to a \emph{canonical}
model $\StdModel{}$ of the theory $\rewtheoryOf{\cR}$ describing computations with $\cR$.
The \emph{choice} of such a model is essential to appropriately understand the property expressed by the formula.
\begin{example}
\label{ExPEvenZeroOdd_GroundVSNonGroundSemanticProperties}
Sentence (\ref{ExPEvenZeroOdd_TermsEitherPositiveOddZero}) is satisfied by the usual 
\emph{least Herbrand model} 
of $\rewtheoryOf{\cR}$  
for $\cR$ in Example \ref{ExPEvenZeroOdd} (as $\rewtheoryOf{\cR}$ can be seen as a set of Horn clauses, see Figure \ref{FigExPEvenZeroOdd_Th} in  Section \ref{SecFirstOrderTheoryOfASimpleInferenceSystem}), thus fitting the intuitive meaning of the sentence. 
But also $\neg(\ref{ExPEvenZeroOdd_TermsEitherPositiveOddZero})$ is satisfied by Clark's \emph{non-ground} \emph{least Herbrand model} discussed below, thus disproving the property if $x$ is instantiated to non-ground terms  (see Example \ref{ExPEvenZeroOdd_TermsEitherPositiveOddZero_NonGroundDisproved} in Section \ref{SecUsingSatisfiabilityInArbitraryInterpretations}).
\end{example}
This paper investigates the use first-order logic methods, techniques, and tools in the analysis of properties of computational systems defined by means of an \ElementaryIS{} so that appropriate solutions to problems like the aforementioned ones can be obtained.
Section \ref{SecPreliminaries} provides some preliminary definitions;
in particular, we remind \emph{Generalized Term Rewriting System}
(\gtrs{} \cite{Lucas_LocalConfluenceOfConditionalAndGeneralizedTermRewritingSystems_JLAMP24}) 
which we often use to illustrate our techniques.
In Section \ref{SecElementaryISs}, borrowing the structure of Smullyan's \emph{Elementary Formal Systems} \cite{Smullyan_TheoryOfFormalSystems_1961}, but using Gentzen's notation for inference rules and deductions \cite{Gentzen_UntersuchungenUberDasLogischeSchliessenPart1_MZ35,Prawitz_NaturalDeductionAProofTheoreticalStudy_1965}, 
we consider inference systems $\GLinference$ consisting of inference rules $\frac{B_1\cdots B_n}{B}$, where $B,B_1,\ldots,B_n$ are \emph{atoms} for some $n\geq 0$, which we call \emph{Elementary Inference Systems}
(\ElementaryIS{s}).
As in \cite{Smullyan_TheoryOfFormalSystems_1961}, 
relations on terms defined by such inference systems are represented 
by predicate symbols $P$ and  obtained by proving atoms 
$P(t_1,\ldots,t_n)$ in 
$\GLinference$ by building appropriate \emph{formula-trees} with 
root $P(t_1,\ldots,t_n)$ (written $\proofInISof{\GLinference}{P(t_1,\ldots,t_n)}$).
A (Horn) first-order theory $\GLtheoryOf{\GLinference}$ is given to 
$\GLinference$ so that provable atoms $A$ in $\GLinference$
are characterized as logical consequences of $\GLtheoryOf{\GLinference}$.
In Section \ref{SecModelsOfTheTheoryOfAnEIS} 
several canonical models are given to
$\GLtheoryOf{\GLinference}$ so that, in Section \ref{SecProvingSemanticPropertiesOfEIS},
properties $\GLformula$ expressed as first-order sentences are said to be 
\emph{semantic properties} of a computational system described by $\GLinference$ 
\emph{relative to a canonical model $\StdModel{}$ of $\GLtheoryOf{\GLinference}$} 
(or just \emph{$\StdModel{}$-properties} of $\GLinference$) if $\GLformula$ is
\emph{satisfied} by $\StdModel{}$, i.e., $\satisfactionInInterpretationOf{\StdModel{}}{\GLformula}$ holds.
We show how to prove and disprove semantic properties in practice.
Section \ref{SecRelatedWork} discusses related work.
Section \ref{SecConclusions} concludes.

\section{Preliminaries}\label{SecPreliminaries}

In the following, we often write \emph{iff} instead of \emph{if and only if}.
We assume some familiarity with the basic notions of term rewriting \cite{BaaNip_TermRewritingAndAllThat_1998,Ohlebusch_AdvancedTopicsInTermRewriting_2002,Terese_TermRewritingSystems_2003}
and first-order logic \cite{Fitting_FirstOrderLogicAndAutomatedTheoremProving_1997,Mendelson_IntroductionToMathematicalLogicFourtEd_1997}.

Given a binary relation $\genrelation\:\subseteq A\times A$ on a set $A$, 
we often write $a\:\genrelation\:b$
instead of $(a,b)\in\:\genrelation$.
The \emph{transitive} closure of $\genrelation$ is denoted by
$\genrelation^+$, and 
its \emph{reflexive and transitive} closure by $\genrelation^*$. 
An element $a\in A$ is \emph{reducible} 
if there exists $b$ such that $a\:\genrelation\:b$. 
In this paper, $\Variables$ denotes a
countable set of \emph{variables} and $\Symbols$ denotes
a \emph{signature of function symbols}, i.e., a set of \emph{function symbols}
$\{f, g, \ldots \}$, each with a fixed \emph{arity} given by a
mapping $ar:\Symbols\rightarrow \mathbb{N}$. The set of
terms built from $\Symbols$ and $\Variables$ is $\Terms$;
and $\GTerms$ is the set of \emph{ground} terms, i.e., without variable occurrences.
The set of variables occurring in $t$ is $\Var(t)$.
We also consider  
signatures of
\emph{predicates} $\SPredicates$. 
Given a signature $\Symbols$, 
a \emph{replacement map} is a mapping 
$\mu$ from symbols in $\Symbols$ to sets of positive numbers
satisfying  $\mu(f)\subseteq \{1,\ldots,ar(f)\}$ for all $f\in\Symbols$
 \cite{Lucas_ContextSensitiveRewriting_CSUR20}.

\subsection{First-order logic}

Given a signature $\Symbols$ of \emph{function symbols}
and a signature $\SPredicates$ of \emph{predicate symbols},
atoms $\GLatom\in\atomsOn{\Symbols,\SPredicates,\Variables}$ and 
first-order formulas  $\GLformula\in\formulasOn{\Symbols,\SPredicates,\Variables}$ on such sets of 
function and predicate symbols with variables in $\Variables$ 
are built in the usual way.
A (definite) Horn clause  (with label $\alpha$) is written $\alpha:A\IF A_1,\ldots,A_n$, 
for atoms $A,A_1,\ldots,A_n$;
if $n=0$, then $\alpha$ is written $A$ rather than $A\IF$.
A first-order theory (FO-theory for short) $\GLtheory$ is a set of sentences
(formulas whose variables are all \emph{quantified}).
An $\Symbols,\SPredicates$-\emph{structure} $\SStructure$ 
(or just \emph{structure} if  no confusion arises)
 consists of a \emph{non-empty}
set $\domainOf{\SStructure}$, called \emph{domain}
 and often denoted $\SemDomain$ if no confusion arises,
 together with an
interpretation  of symbols $f\in\Symbols$ and $P\in\SPredicates$  
as mappings $f^\SStructure$ and
relations $P^\SStructure$ on $\SemDomain$, respectively. 
Then, the usual interpretation of first-order formulas
with respect to $\SStructure$  is considered \cite[page 60]{Mendelson_IntroductionToMathematicalLogicFourtEd_1997}. 
An $\Symbols,\SPredicates$-model for a theory $\GLtheory$
is just a structure $\SStructure$ that makes all the sentences of the theory true, written $\SStructure\models\GLtheory$.
A theory $\GLtheory$ that has a model is said to be \emph{consistent}.
Two theories are \emph{equivalent} if they have the same models.
A formula $\GLformula$ is a \emph{logical consequence} of a theory $\GLtheory$ (written $\GLtheory\models\GLformula$) 
iff every model $\SStructure$
of $\GLtheory$ is also a model of $\GLformula$.
Also, $\deductionInThOf{\GLtheory}{\GLformula}$ means that $\GLformula$ is \emph{deducible} from 
$\GLtheory$ by using 
a correct and complete deduction procedure.

\subsection{Generalized Term Rewriting Systems}
A \emph{Generalized Term Rewriting System (\gtrs{} \cite[Section 7]{Lucas_LocalConfluenceOfConditionalAndGeneralizedTermRewritingSystems_JLAMP24})} is a tuple
$\cR=(\Symbols,\SPredicates,\mu,H,R)$ where 
$\Symbols$ is a signature of \emph{function symbols},
$\SPredicates$ is a  signature of \emph{predicate symbols}, including 
at least $\rew{}$ and $\rews{}$, 
$\mu\in\Rmaps{\Symbols}$, 
 $H$ is a (possibly empty) set of clauses $A\IF\gencond$, 
 where $\rootTerm(A)\notin\{\rew{},\rews{}\}$, and 
$R$ is a set of rewrite rules $\lhsr\to\rhsr\IF\gencond$ such that $\lhsr\notin\Variables$. 
In both cases, $\gencond$ is a sequence of atoms.
Note that rules in $R$ are Horn clauses.

\section{Elementary Inference Systems}
\label{SecElementaryISs}

In this paper, we consider the following class of inference systems.

\begin{definition}[Elementary inference system]
\label{DefInferenceSystem}
Let $\Symbols$ and $\SPredicates$ be signatures of function  and predicate symbols, respectively,
and $\Variables$ be a set of variables.
An inference rule $\rho:\frac{B_1\cdots B_n}{B}$ (with label $\rho$) 
is called  \emph{elementary} if
$B,B_1,\ldots,B_n\in\atomsOn{\Symbols,\SPredicates,\Variables}$ are \emph{atoms}.
An \emph{elementary} inference system (\ElementaryIS{} for short) is
a tuple $\GLinference=(\Symbols,\SPredicates,\SetOfInferenceRules)$, where $\SetOfInferenceRules$ 
is a set of elementary inference rules.
\end{definition}
\begin{remark}
\label{RemISBeyondEIS}
In the literature, inference rules may have a more elaborated structure, typically using \emph{sequents} (usually written $\Gamma\vdash\GLformula$, where $\Gamma$ is an ``environment'', typically giving values to variables occurring in $\GLformula$, which is an arbitrary formula) instead of just atoms $\GLatom$ as components of the rule, see, e.g., \cite[Section 2.1]{Kahn_NaturalSemantics_STACS87}.
The structural simplicity of \ElementaryIS{s} is important to obtain also simple definitions of provability, etc.
\end{remark}
Given an \ElementaryIS{} $\GLinference=(\Symbols,\SPredicates,\SetOfInferenceRules)$, we often write $\rho\in\GLinference$ instead of $\rho\in\SetOfInferenceRules$.
\begin{figure}[t]
\begin{tabular}{c@{\qquad\qquad\quad}c}
\begin{minipage}{4cm}
\begin{tabular}{l@{\quad}c}
$(\RuleReflexivity)$ & $\bigfracLabel{}{x \rightarrow^{\ast} x}$  
\\[0.6cm] 
$(\RuleCompatibility)$ & $\bigfracLabel{x \rightarrow y\qquad y
  \rightarrow^{*} z}{x  
\rightarrow^{*} z}$ 
\end{tabular}
\end{minipage}
&
\begin{minipage}{7cm}
\begin{tabular}{l@{\quad}c}
$(\RulePropagation)_{f,i}$ & $ \bigfracLabel{x_i \rightarrow y_i}
{f(x_{1},\ldots,x_{i},\ldots,x_{k}){}\rightarrow{}f(x_{1},\ldots,y_{i},\ldots,x_{k})}$ 
\\[0.6cm]
$(\RuleHornClause)_{B\IFhorn B_1,\ldots,B_n}$ &
$\bigfracLabel{B_1  \quad \cdots \quad B_n}{B}
$ 
\end{tabular}
\end{minipage}
\end{tabular}
\caption{Generic elementary inference rules for a \gtrs{}}
\label{FigElementaryInferenceSystemGTRS}
\end{figure}
\begin{definition}[\ElementaryIS{} of a \gtrs]
\label{DefEISofAGTRS}
The \ElementaryIS{}
$\GLinferenceOf{\cR}=(\Symbols,\SPredicates,\SetOfInferenceRules)$ of
a \gtrs{} $\cR=(\Symbols,\SPredicates,\mu,H,R)$ is
(using the generic inference rules in Figure \ref{FigElementaryInferenceSystemGTRS}):
\[\SetOfInferenceRules = \{(\RuleReflexivity),(\RuleCompatibility)\}\cup\bigcup_{f\in\Symbols,i\in\mu(f)}\{(\RulePropagation)_{f,i}\}
\cup\bigcup_{\alpha\in H\cup R}\{(\RuleHornClause)_\alpha\}\]
\end{definition}

\subsection{Proofs with Elementary Inference Systems}

A \emph{finite proof tree} $T$ in  $\GLinference$
with root $\GLgoal\in\atomsOn{\Symbols,\SPredicates,\Variables}$ 
is either:
\begin{itemize}
\item 
an \emph{open goal}, simply denoted as $\GLgoal$; 
or
\item  
a \emph{derivation tree} 
denoted as
$
\frac{T_1 \quad \cdots \quad T_n}{\GLgoal}(\rho)
$,
where 
$T_1$,\ldots,$T_n$ are finite proof trees in $\GLinference$ (for $n\geq 0$; if $n=0$ instead of 
$\frac{}{G}(\rho)$
we just write
$\ol{G}$), and 
$\rho : \frac{B_{1} \cdots  B_{n}}{B}\in\GLinference$ is an inference rule
such that
$\GLgoal = \sigma(B)$, and
$\rootTree(T_1)=\sigma(B_{1}), \ldots, \rootTree(T_n)=\sigma(B_{n})$
for some substitution $\sigma$.
\end{itemize}
Note 
that 
inference rules $\frac{B_{1} \cdots  B_{n}}{B}$
in $\GLinference$ are viewed
as \emph{schemes of rules} whose head $B$ should \emph{match} the goal $\GLgoal$
with a matching substitution $\sigma$ (see \cite[Chapter I, \#A, \textsection 2]{Smullyan_TheoryOfFormalSystems_1961}).
A finite proof tree $T$ is \emph{closed} if it 
contains no open goals.
\begin{definition}[Provable atom]\label{DefProbableGoal}
Let $\GLinference$ 
be an \ElementaryIS{}. An atom $\GLatom$ is \emph{provable} in $\GLinference$, written $\proofInISof{\GLinference}{\GLatom}$,
if there is a closed proof tree $T$ with $\rootTree(T)=\GLatom$ using $\GLinference$.
\end{definition}

\begin{remark}
\label{RemProofsInISBeyondProofsInEIS}
In the literature, proofs with inference rules may have a more elaborated definition.
For instance, the usual rule dealing with the assignment instruction of imperative languages, see, e.g., \cite[page 46]{Plotkin_AStructuralApproachToOperationalSemantics_JLAP04}:
\begin{eqnarray}
\bigfrac{\langle e,\varsigma\rangle\longrightarrow^*\langle m,\varsigma\rangle}{\langle v:=e,\varsigma\rangle\longrightarrow\varsigma[v\mapsto m]}
\label{LblAssignmentRuleSTS}
\end{eqnarray}
where 
$e$ is an expression, 
$\varsigma$ is a \emph{store}, i.e., a mapping from variables to numbers, 
$m$ is a number, 
$v$ is a program variable,
and 
$\varsigma[v\mapsto m]$ is a new store obtained from $\varsigma$ so that variable $v$  is bounded to $m$ in $\varsigma[v\mapsto m]$, and any other variable $v'$ different from $v$ remains bounded in $\varsigma[v\mapsto m]$ as it was in $\varsigma$ (see \cite[Section 2.1]{Plotkin_AStructuralApproachToOperationalSemantics_JLAP04} for the technical details).
The \emph{update} $\varsigma[v\mapsto m]$ of a store $\varsigma$ using (\ref{LblAssignmentRuleSTS})
cannot be handled as the application of a substitution 
as required by Definition \ref{DefProbableGoal}.
However, if we assume finitely many program variables $v_1,\ldots,v_k$, rule 
(\ref{LblAssignmentRuleSTS}) 
could be seen as $k$ elementary rules as follows:
\begin{eqnarray*}
\bigfrac{\langle e,\fS{st}(m_1,\ldots,m_i,\ldots,m_k)\rangle\longrightarrow^*\langle m,\fS{st}(m_1,\ldots,m_i,\ldots,m_k)\rangle}{\langle v_i:=e,\varsigma\rangle\longrightarrow\fS{st}(m_1,\ldots,m,\ldots,m_k)}
\end{eqnarray*}
where $m,m_1,\ldots,m_k$ are variables (disjoint from $v_1,\ldots,v_k$).
However, $e$ should be written using indexed variables $v_i$.
Furthermore, evaluation rules for variables should also be decomposed into $k$ rules as follows:
\begin{eqnarray*}
\langle v_i,st(m_1,\ldots,m_i,\ldots,m_k)\rangle\longrightarrow \langle m_i,st(m_1,\ldots,m_i,\ldots,m_k)\rangle
\end{eqnarray*}
instead of the (single) Variable rule in \cite[page 42]{Plotkin_AStructuralApproachToOperationalSemantics_JLAP04}, i.e.,
\begin{eqnarray*}
\langle v,\varsigma\rangle\longrightarrow\langle\varsigma(v),\varsigma\rangle
\end{eqnarray*}
\end{remark}
For each $n$-ary predicate $P\in\SPredicates$,
the relation on terms $P^\GLinference$ 
defined by $\GLinference$ for $P$ is
\begin{eqnarray*}
P^\GLinference & = & \{P(t_1,\ldots,t_n)\mid t_1,\ldots,t_n\in\Terms,\proofInISof{\GLinference}{P(t_1,\ldots,t_n)}\}\label{DefRelationDefinedByIS}
\end{eqnarray*}
Provability of (atomic) goals in an \ElementaryIS{} is obviously 
\emph{preserved} under substitution application.

\begin{proposition}
\label{PropStabilityOfProvabilityInSimpleIS}
Let $\GLinference$ be an \ElementaryIS, $\GLatom$ be an atom,
and $\sigma$ be a substitution.
If $\proofInISof{\GLinference}{\GLatom}$, then 
$\proofInISof{\GLinference}{\sigma(\GLatom)}$.
\end{proposition}
A finite proof tree $T$ is a \emph{proper prefix} of a finite proof tree $T'$ (written $T \subset T'$) 
if there
  are one or more open goals $\GLgoal_1,\ldots,\GLgoal_n$ in $T$ such that $T'$ is
  obtained from $T$ by replacing each $\GLgoal_i$ by a finite derivation tree
  $T_i$ with root $\GLgoal_i$.
  An \emph{infinite proof tree} $T$ is an infinite increasing chain of
  finite proof trees, i.e., a sequence $(T_i)_{i\in\mathbb{N}}$ such
  that for all $i$, $T_i \subset T_{i+1}$.
Since for all $i\in\naturals$, $\rootTree(T_i)=\rootTree(T_{i+1})$, we write $\rootTree(T)=\rootTree(T_0)$.
A finite proof tree $T$ is \emph{well-formed} if it is either an 
open goal, or a closed proof tree, or a derivation tree  
$
\frac{T_1 \quad \cdots \quad T_n}{\GLgoal}(\rho),
$
where $T_1,\ldots,T_{i-1}$ are closed for some $1\leq i\leq n$,
$T_i$ is a well-formed but not closed finite proof tree, and $T_{i+1},\ldots,T_n$ are open goals.
Note the \emph{left-to-right}  construction of the proof tree.
An infinite proof tree is well-formed if it is an increasing 
chain of well-formed finite proof trees.
As an application of the notion of operational termination \cite{LucMarMes_OperationalTerminationOfConditionalTermRewritingSystems_IPL05} we obtain the following.

\begin{definition}
\label{DefOpTermEIS}
(cf. \cite[Definition 4]{LucMarMes_OperationalTerminationOfConditionalTermRewritingSystems_IPL05})
An \ElementaryIS{} $\GLinference$ is called \emph{operationally terminating}
if no infinite well-formed proof tree for $\GLinference$ exists.
\end{definition}
In \cite{Lucas_LocalConfluenceOfConditionalAndGeneralizedTermRewritingSystems_JLAMP24,Lucas_TerminationOfGeneralizedTermRewritingSystems_FSCD24}, 
no inference system was given to a \gtrs{}. Only  \emph{termination} (of the one-step relation $\rew{\cR}$) is discussed in \cite{Lucas_TerminationOfGeneralizedTermRewritingSystems_FSCD24}.
Using Definition \ref{DefEISofAGTRS}, we introduce the following:

\begin{definition}
\label{DefOpTermGCTRS}
A \gtrs{} $\cR$ is operationally terminating if $\GLinferenceOf{\cR}$ is.
\end{definition}
For \emph{binary} predicates $P\in\SPredicates$,
\emph{termination} of the binary relation on terms $P^\GLinference$ 
is defined as expected:
\begin{definition}
\label{DefTerminationOfBinaryPredicate}
Let $\GLinference=(\Symbols,\SPredicates,\SetOfInferenceRules)$ be an 
\ElementaryIS{} and $P\in\SPredicates$ be a binary predicate.
We say that $P$ is $\GLinference$-terminating if there is no infinite sequence 
$t_1,t_2,\ldots$ of terms $t_i\in\Terms$ such that, for all $i\geq 1$,
$P^\GLinference(t_i,t_{i+1})$ holds.
\end{definition}
For \gtrs{s} $\cR$, termination of $\cR$, i.e., termination of $\rew{\cR}$
in the usual sense \cite[Section 7.5]{Lucas_LocalConfluenceOfConditionalAndGeneralizedTermRewritingSystems_JLAMP24} coincides with termination of $\rew{}^{\GLinferenceOf{\cR}}$ in Definition \ref{DefTerminationOfBinaryPredicate}. 

\subsection{First-Order Theory of an Elementary Inference System}
\label{SecFirstOrderTheoryOfASimpleInferenceSystem}

As done in, e.g., \cite{HanMil_DerivingMixedEvaluationFromStandardEvaluationForASimpleFunctionalLanguage_MPC89,HanMil_FromOperationalSemanticsToAbstractMachinesPreliminaryResults_LFP90},
from each elementary inference rule $\rho:\frac{B_1,\ldots,B_n}{B}$ and $\vec{x}=\Var(B,B_1,\ldots,B_n)$,
we obtain a sentence $\sentenceFrom{\rho}$ 
(which we call a \emph{definite Horn sentence}) as follows
\begin{eqnarray*}
(\forall\vec{x}) & B_1\wedge\cdots\wedge B_n\Rightarrow B\label{LblSentenceFromARule}
\end{eqnarray*}
If $n=0$, we just write $(\forall\vec{x})\:B$;
if $\vec{x}$ is  empty, we just write $B_1\wedge\cdots\wedge B_n\Rightarrow B$.
Given an \ElementaryIS{} $\GLinference$, we obtain a theory 
$\GLtheoryOf{\GLinference}  =  \{\sentenceFrom{\rho}\mid\rho\in\GLinference\}$.
\begin{example}
For $\cR$ in Example \ref{ExPEvenZeroOdd} and $\GLinferenceOf{\cR}$ in Figure \ref{FigExPEvenZeroOdd_IS},
$\rewtheoryOf{\cR}=\GLtheoryOf{\GLinferenceOf{\cR}}$ is displayed in 
Figure \ref{FigExPEvenZeroOdd_Th}.
By abuse of notation, we use $\rho$ instead of $\overline{\rho}$ to denote sentences $\overline{\rho}$  obtained from inference rules $\rho$.
\begin{figure}[t]
\begin{tabular}{c@{}c}
\begin{minipage}{7.5cm}
\[
\begin{array}{crl}
(\RuleReflexivity) 
& (\forall x) & x\rews{}x
\\
(\RuleCompatibility)
& (\forall x,y,z) 
& x\rew{}y\wedge y\rews{}z\Rightarrow x\rews{}z
\\
(\RuleHornClause)_{(\ref{ExPEvenZeroOdd_horn1})}
& (\forall x) 
& x  \geq  0
\\
(\RuleHornClause)_{(\ref{ExPEvenZeroOdd_horn2})}
& (\forall x,y) 
& x\geq y\Rightarrow \fS{s}(x) \geq \fS{s}(y)
\end{array}
\]
\end{minipage}
&
\begin{minipage}{7.5cm}
\[
\begin{array}{crl}
(\RuleHornClause)_{(\ref{ExPEvenZeroOdd_horn3})}
& (\forall x) 
& x\rews{}\fS{s}(\fS{s}(\fS{0}))\Rightarrow \fS{peven}(x)
\\
(\RuleHornClause)_{(\ref{ExPEvenZeroOdd_horn4})}
& (\forall x) 
& x\rews{}\fS{s}(\fS{0})\Rightarrow \fS{odd}(x)
\\
(\RuleHornClause)_{(\ref{ExPEvenZeroOdd_horn5})}
& (\forall x) 
& x\rews{}\fS{0}\Rightarrow \fS{zero}(x)
\\
(\RuleHornClause)_{(\ref{ExPEvenZeroOdd_rule1})}
& (\forall x) 
& x\geq\fS{s}(\fS{0})\Rightarrow \fS{s}(\fS{s}(x)) \to   x
\end{array}
\]
\end{minipage}
\end{tabular}
\caption{Theory $\rewtheoryOf{\cR}$ for $\cR$ in Example \ref{ExPEvenZeroOdd}}
\label{FigExPEvenZeroOdd_Th}
\end{figure}
\end{example}
The following result establishes the equivalence between provability of atoms $\GLatom$ in
an \ElementaryIS{} $\GLinference$ and deduction of $\GLatom$ (i.e., the universal closure of $\GLatom$)
in $\GLtheoryOf{\GLinference}$.

\begin{proposition}\label{PropProofInEISsAsDeduction}
Let 
$\GLinference$ be an \ElementaryIS{} and $\GLatom$ be an atom
with variables 
$\vec{x}$.
Then,
$\proofInISof{\GLinference}{\GLatom}$ iff 
$\deductionInThOf{\GLtheoryOf{\GLinference}}{(\forall\vec{x})\:\GLatom}$.
\end{proposition}
\begin{remark}[Provability for \gtrs{s} $\cR$]
Since $\GLtheoryOf{\GLinferenceOf{\cR}}$ and the theory $\rewtheoryOf{\cR}$ associated to $\cR$ in \cite[Definition 52]{Lucas_LocalConfluenceOfConditionalAndGeneralizedTermRewritingSystems_JLAMP24} coincide, 
Proposition \ref{PropProofInEISsAsDeduction} shows that defining rewriting steps $s\rew{\cR}t$ as deduction of $s\rew{}t$ (i.e., $(\forall\vec{x})\:s\rew{}t$) in 
 $\rewtheoryOf{\cR}$ \cite[Section 7.5 \& Definition 8]{Lucas_LocalConfluenceOfConditionalAndGeneralizedTermRewritingSystems_JLAMP24} is \emph{equivalent} to provability of $s\rew{}t$ in $\GLinferenceOf{\cR}$.
 \end{remark}
In the following, for \gtrs{s} we use  $\rewtheoryOf{\cR}$ rather than $\GLtheoryOf{\GLinferenceOf{\cR}}$.

\section{Models of Elementary Inference Systems}
\label{SecModelsOfTheTheoryOfAnEIS}

Every FO-sentence $\GLformula$ can be expressed as a set  
$C_\GLformula$ of clauses (a \emph{standard} form of $\GLformula$ 
\cite[Section 4.2]{ChaLee_symbolicLogicAndMechanicalTheoremProving_1973})
so that \emph{$C_\GLformula$ is inconsistent iff $\GLformula$ is} \cite[Theorem 4.1]{ChaLee_symbolicLogicAndMechanicalTheoremProving_1973}.
However, due to \emph{skolemization}, 
$\GLformula$ and $C_\GLformula$ are, in general,  \emph{not} equivalent.
\begin{example}
\label{ExSkolemizationAndLogicalEquivalence}
The set  
$C_\GLformula=\{\fS{P}(\fS{a})\}$ is a standard form of 
$\GLformula=(\exists x)\fS{P}(x)$.
The interpretation $\SStructure$ with domain $\SStructure=\{1,2\}$, 
$\fS{a}^\SStructure=1$, and $\fS{P}^\SStructure=\{(2)\}$ is a model of $\GLformula$ but it is \emph{not} a model of $C_\GLformula$ \cite[page 49]{ChaLee_symbolicLogicAndMechanicalTheoremProving_1973}.
\end{example}
Dealing with sets of clauses, we usually consider \emph{Herbrand interpretations}.

\subsection{Herbrand interpretations.}
The domain  of an Herbrand 
$\Symbols,\SPredicates$-interpretation
$\HerbrandInterpretation$ 
(or just $\fS{H}$-in\-ter\-pre\-ta\-tion, if no confusion arises) 
is $\domainOf{\HerbrandInterpretation}=\GTerms$, which, by the non-emptiness requirement on interpretations (see Section \ref{SecPreliminaries}), must be \emph{non-empty}; hence $\Symbols$ must contain at least 
 one constant.
Each $k$-ary
function symbol $f\in\Symbols$ is given a mapping $f^\SStructure:\GTerms\times\cdots\GTerms\to\GTerms$ defined by $f^\SStructure(t_1,\ldots,t_k)=f(t_1,\ldots,t_k)$ for all $t_1,\ldots,t_k\in\GTerms$.
Since the domain and function symbol interpretation are fixed, 
$\HerbrandInterpretation$ is usually described/identified as
a subset $\HerbrandInterpretation\subseteq\HerbrandBase$ of \emph{ground} atoms in the Herbrand Base 
$\HerbrandBase=\atomsOn{\Symbols,\SPredicates,\emptyset}$
\cite{ChaLee_symbolicLogicAndMechanicalTheoremProving_1973}.
 Then, $n$-ary predicates $P\in\SPredicates$ are interpreted by 
 $P^\SStructure=\{(t_1,\ldots,t_n)\in\GTerms^n\mid P(t_1,\ldots,t_n)\in\HerbrandInterpretation\}$
 \cite[page 53]{ChaLee_symbolicLogicAndMechanicalTheoremProving_1973}.
  
A set of clauses is unsatisfiable (i.e., inconsistent) iff it has no Herbrand model
\cite[Theorem 4.2]{ChaLee_symbolicLogicAndMechanicalTheoremProving_1973}.
This may \emph{fail} to hold for arbitrary theories.
\begin{example}
 \label{ExConsistencyAndHerbrandInterpretations}
Note that $\GLtheory=\{\fS{P}(\fS{a}), (\exists x) \neg\fS{P}(x)\}$ 
is \emph{not} a set of clauses due to the existential quantification of the second formula. It is satisfied by $\SStructure$
 with domain $\SStructure=\{0,1\}$, $\fS{a}^\SStructure=0$ and 
 $\fS{P}^\SStructure=\{(0)\}$ but none of the two possible Herbrand interpretations $\HerbrandInterpretation_1=\emptyset$ and $\HerbrandInterpretation_2=\{\fS{P}(\fS{a})\}$
satisfies $S$ \cite[pp.\ 17--18]{Lloyd_FoundationsOfLogicProgramming_1987}.
 \end{example}
 This motivates the following.
 
 \begin{definition}[H-consistency]
 \label{DefHConsistency}
 A theory $\GLtheory$ is \emph{H-consistent} if it has a Herbrand model.
 Otherwise, it is \emph{H-inconsistent}.
 \end{definition}
 H-consistent theories are consistent, but not vice versa, as Example \ref{ExConsistencyAndHerbrandInterpretations} shows.
Furthermore, in sharp contrast to inconsistency, H-inconsistency is \emph{not} preserved by 
 standarization of formulas.
 
 \begin{example}
 \label{ExSkolemizationAndHinconsistency}
 Remind that $\GLtheory=\{\fS{P}(\fS{a}), (\exists x) \neg\fS{P}(x)\}$ in Example \ref{ExConsistencyAndHerbrandInterpretations} is 
  H-inconsistent.
However, $C_\GLtheory=\{\fS{P}(\fS{a}), \neg\fS{P}(c)\}$, where $c$ is a fresh (Skolem) 
constant, 
is a standard version of $\GLtheory$ which is H-consistent as 
the H-interpretation $\HerbrandInterpretation=\{\fS{P}(\fS{a})\}$
is a model of $C_\GLtheory$.
\end{example}
 The standard semantics for 
sets 
of definite Horn clauses 
over signatures $\Symbols$ and $\SPredicates$ of function and predicate symbols
(where $\Symbols$ contains at least one constant), 
using variables in $\Variables$ \cite{EmdKow_TheSemanticsOfPredicateLogicAsAProgrammingLanguage_JACM76}
considers  
\emph{Herbrand $\Symbols,\SPredicates$-interpretations}
$\HerbrandInterpretation$ viewed as subsets 
$\HerbrandInterpretation\subseteq\HerbrandBaseOf{\Symbols,\SPredicates}=\atomsOn{\Symbols,\SPredicates,\emptyset}$ of \emph{ground} atoms.
 We apply these ideas to $\ElementaryIS{s}$ through $\GLtheoryOf{\GLinference}$, 
 which is a set of definite Horn clauses.
 
 \subsection{Least Herbrand model of an \ElementaryIS} 
Every set 
 of definite Horn clauses has a \emph{least} (with respect to set inclusion) 
 Herbrand 
 $\Symbols,\SPredicates$-model (of ground atomic consequences)
  \cite[Section 5]{EmdKow_TheSemanticsOfPredicateLogicAsAProgrammingLanguage_JACM76}. 

 \begin{definition}[Canonical Herbrand Model of an \ElementaryIS]
 \label{DefCanonicalHerbrandModelOfAnEIS}
 Let $\GLinference=(\Symbols,\SPredicates,I)$ be an \ElementaryIS{}.
The canonical $\fS{H}$-model of $\GLinference$ is:
 \[
 \StdModelOf{\GLinference}
 =\{\GLatom\in\HerbrandBaseOf{\Symbols,\SPredicates}\mid \GLtheoryOf{\GLinference}\models\GLatom\}
 =\{\GLatom\in\HerbrandBaseOf{\Symbols,\SPredicates}\mid \GLtheoryOf{\GLinference}\vdash\GLatom\}=\{\GLatom\in\HerbrandBaseOf{\Symbols,\SPredicates}\mid\proofInISof{\GLinference}{\GLatom}\}
 \] 
 \end{definition}
Proposition \ref{PropProofInEISsAsDeduction} justifies the last equality.
As in \cite{EmdKow_TheSemanticsOfPredicateLogicAsAProgrammingLanguage_JACM76}, the \emph{canonicity} of $\StdModelOf{\GLinference}$ comes from the fact that 
every atom in $\StdModelOf{\GLinference}$ belongs to \emph{every}  $\fS{H}$-model of
$\GLtheoryOf{\GLinference}$.

\subsection{Least V-Herbrand model of an \ElementaryIS} 

Clark extended van Emden and Kowalski's approach to \emph{non-ground (but also called Herbrand) interpretations}
$\VHerbrandInterpretation$ whose interpretation domain is $\Terms$, 
rather than $\GTerms$,
$k$-ary function symbols $f\in\Symbols$ are given mappings
$f^{\VHerbrandInterpretation}:\Terms^k\to\Terms$,
and the interpretation of predicate symbols is usually represented as a subset $\VHerbrandInterpretation\subseteq\VHerbrandBaseOf{\Symbols,\SPredicates,\Variables}$ 
of the non-ground Herbrand base $\VHerbrandBaseOf{\Symbols,\SPredicates,\Variables}=
\atomsOn{\Symbols,\SPredicates,\Variables}$
(or just $\VHerbrandBase$ if no confusion arises) 
consisting of all atoms (possibly with variables).
We call them V-Herbrand $\Symbols,\SPredicates$-interpretations, or just
$\widehat{\fS{H}}$-interpretations.
Note that, since $\Terms$ is never empty due to the non-emptiness of 
$\Variables$, we do not need to impose that $\Symbols$ contains a constant symbol.
As for the standard case, Clark 
shows the existence of
a \emph{least}  (with respect to set inclusion) $\widehat{\fS{H}}$-model 
\cite[Theorem 3.6]{Clark_PredicateLogicAsAComputationalFormalism_TR79}.
Accordingly, we introduce the following.

\begin{definition}[Canonical V-Herbrand Model of an \ElementaryIS]
 \label{DefCanonicalVHerbrandModelOfAnEIS}
 Let $\GLinference=(\Symbols,\SPredicates,I)$ be an \ElementaryIS{}.
The canonical $\widehat{\fS{H}}$-model of $\GLinference$ is:
 \[\begin{array}{rcl}
 \StdVModelOf{\GLinference} 
 & = & \{\GLatom\in\VHerbrandBaseOf{\Symbols,\SPredicates,\Variables}\mid \GLtheoryOf{\GLinference}\models(\forall\vec{x})\GLatom\}
 =\{\GLatom\in\VHerbrandBaseOf{\Symbols,\SPredicates,\Variables}\mid \deductionInThOf{\GLtheoryOf{\GLinference}}{(\forall\vec{x})\GLatom}\}\\
 & = & \{\GLatom\in\VHerbrandBaseOf{\Symbols,\SPredicates,\Variables}\mid\proofInISof{\GLinference}{\GLatom}\}
 \end{array}
 \] 
 \end{definition}
Such a model can be considered as the \emph{canonical} model of the 
\emph{non-ground} model-theoretic semantics of $\GLinference$.
Note that  $\StdModelOf{\GLinference}\subseteq \StdVModelOf{\GLinference}$.
As we will see in Section \ref{SecProvingSemanticPropertiesOfEIS}, having different \emph{canonical} 
models is essential to define different kind of properties.

For \gtrs{s} $\cR$, we write 
$\StdModelOf{\cR}$ and $\StdVModelOf{\cR}$ rather than
$\StdModelOf{\GLinferenceOf{\cR}}$ and $\StdVModelOf{\GLinferenceOf{\cR}}$.
The most natural model for \gtrs{s} is $\StdVModelOf{\cR}$ as the interpretation domain consists of arbitrary (not only ground) terms, which are the
usual `subject' expressions in term rewriting.
However, 
$\StdModelOf{\cR}$
captures important properties as well (see Example \ref{ExPEvenZeroOdd_GroundVSNonGroundSemanticProperties}).

\subsection{Grounding the least V-Herbrand model}
\label{SecGroundingTheLeastVHerbrandModel}

Let $\Symbols,\Symbols'$ and $\SPredicates,\SPredicates'$ be signatures of function and predicate symbols such that $\Symbols\subseteq\Symbols'$ and $\SPredicates\subseteq\SPredicates'$.
It is clear that every $\Symbols',\SPredicates'$-structure $\SStructure$ 
can be seen as a 
$\Symbols,\SPredicates$-structure $\restrictTo{\SStructure}{\Symbols,\SPredicates}$
with the \emph{same} domain of interpretation $\domainOf{\SStructure}$
and taking from $\SStructure$ the interpretations
$f^\SStructure$ and $P^\SStructure$ for all $f\in\Symbols$ and $P\in\SPredicates$.
In the following, we often silently use $\Symbols',\SPredicates'$-structure as an 
$\Symbols,\SPredicates$-structure by assuming the previous adaptation.

Let $\Symbols$ be a signature and $\Variables$ be a denumerable, infinite set of variables such that $\Symbols\cap\Variables=\emptyset$.
Since variables in subject terms $t$ behave like constant symbols in any rewriting sequence, as in, e.g., \cite[page 224]{AveLor_ConditionalRewriteSystemsWithExtraVariablesAndDeterministicLogicPrograms_LPAR94} and \cite[page 78]{BaaNip_TermRewritingAndAllThat_1998}, given a term
$t$, a term  $t^\downarrow$ is
obtained by replacing each occurrence of $x\in\Variables$ in $t$ by 
a fresh constant $c_x\notin\Symbols\cup\Variables$.
We let $\GroundedVariables=\{c_x\mid x\in\Variables\}$ and 
$\GroundedSymbols=\Symbols\cup \GroundedVariables$.
Given a term $t\in\Terms$, its \emph{grounded} version is $\grounding{t}\in\GroundedTerms$.
Vice versa: given $t\in\GroundedTerms$, its \emph{ungrounded} version $t^\uparrow\in\Terms$ is obtained by
replacing, for all $x\in\Variables$, each constant $c_x$ in $t$ by $x$.
For all terms $t\in\Terms$, $(t^\downarrow)^\uparrow=t$; and 
for all terms $t\in\GroundedTerms$, $(t^\uparrow)^\downarrow=t$. 
Also, given $\GLatom\in\atomsOn{\Symbols,\SPredicates,\Variables}$, $\grounding{\GLatom}\in\atomsOn{\GroundedSymbols,\SPredicates,\emptyset}$ is its \emph{grounded} version; given $\GLatom\in\atomsOn{\GroundedSymbols,\SPredicates,\emptyset}$, 
$\ungrounding{\GLatom}\in\atomsOn{\Symbols,\SPredicates,\Variables}$ is its \emph{ungrounded} version.
Given a substitution $\sigma=\{x_1\mapsto t_1,\ldots,x_n\mapsto t_n\}$, we let 
$\sigma^\downarrow=\{x_1\mapsto t^\downarrow_1,\ldots,x_n\mapsto t^\downarrow_n\}$.
Grounding of variables preserves pattern matching
in the following sense.

\begin{proposition}\label{PropGroundingPreservesMatching}
Let $p,t\in\Terms$, $A,B\in\atomsOn{\Symbols,\SPredicates,\Variables}$, 
and $\sigma$ be a substitution. 
Then, (i) $t=\sigma(p)$ iff
$\grounding{t}=\grounding{\sigma}(p)$
and
(ii) 
$A=\sigma(B)$ iff $\grounding{A}=\grounding{\sigma}(B)$.
\end{proposition}
As a consequence of Proposition \ref{PropGroundingPreservesMatching}
and the definition of provability in an \ElementaryIS, we have the following.

\begin{proposition}\label{PropProofInEISsOfAtomsAndGroundedAtoms}
Let 
$\GLinference=(\Symbols,\SPredicates,\SetOfInferenceRules)$ be an \ElementaryIS{} and $\GLatom\in\atomsOn{\Symbols,\SPredicates,\Variables}$.
Then,
$\proofInISof{\GLinference}{\GLatom}$ iff 
$\proofInISof{\GLinference}{\grounding{\GLatom}}$.
\end{proposition}
By Proposition \ref{PropProofInEISsOfAtomsAndGroundedAtoms}, 
V-Herbrand $\Symbols,\SPredicates$-interpretations $\VHerbrandInterpretation\subseteq\atomsOn{\Symbols,\SPredicates,\Variables}$, 
can be  \emph{grounded} into an `equivalent'
Herbrand $\GroundedSymbols,\SPredicates$-interpretation 
$\grounding{\VHerbrandInterpretation}=\{\grounding{\GLatom}\mid \GLatom\in\VHerbrandInterpretation\}\subseteq\atomsOn{\GroundedSymbols,\SPredicates,\emptyset}$, which we often call an $\grounding{\fS{H}}$-interpretation if no confusion arises.
\begin{definition}\label{DefGroundedCanonicalModel}
The \emph{grounded canonical $\grounding{\fS{H}}$-model} 
$\StdGroundedVModelOf{\GLinference}$ of $\GLinference$ is $\grounding{\StdVModelOf{\GLinference}}$.
\end{definition}
Given an \ElementaryIS{} $\GLinference=(\Symbols,\SPredicates,\SetOfInferenceRules)$, 
$\StdGroundedVModelOf{\GLinference}$ (viewed as an $\Symbols,\SPredicates$-interpretation) is a model of $\GLtheoryOf{\GLinference}\subseteq\formulasOn{\Symbols,\SPredicates,\Variables}$.
\begin{theorem}\label{TheoStdModelOfSimpleISatisfiesFOtheoryOfIS}
Let $\GLinference=(\Symbols,\SPredicates,\SetOfInferenceRules)$ be an $\ElementaryIS$.
Then, $\satisfactionInThOf{\StdGroundedVModelOf{\GLinference}}{\GLtheoryOf{\GLinference}}$.
\end{theorem}
According to 
\cite[page 39]{Hodges_AShorterModelTheory_1997}, two 
$\Symbols,\SPredicates$-structures are \emph{equivalent} if
they satisfy the same formulas $\GLformula\in\formulasOn{\Symbols,\SPredicates,\Variables}$.
Then,
$\StdVModelOf{\GLinference}$ and $\StdGroundedVModelOf{\GLinference}$ are \emph{equivalent}: 
\begin{theorem}
\label{TheoEquivalenceOfVModelAndGroundedVModel}
Let $\GLinference=(\Symbols,\SPredicates,\SetOfInferenceRules)$ be an \ElementaryIS{} and $\GLformula\in\formulasOn{\Symbols,\SPredicates,\Variables}$.
Then, $\satisfactionInThOf{\StdVModelOf{\GLinference}}{\GLformula}$ iff $\satisfactionInThOf{\StdGroundedVModelOf{\GLinference}}{\GLformula}$.
\end{theorem}
Theorem \ref{TheoEquivalenceOfVModelAndGroundedVModel} justifies that $\StdGroundedVModelOf{\GLinference}$ 
is called `canonical' in Definition \ref{DefGroundedCanonicalModel},
as it is equivalent (on formulas $\GLformula\in\formulasOn{\Symbols,\SPredicates,\Variables}$) to the canonical model 
$\StdVModelOf{\GLinference}$.
We also have the following ``quantifier elimination'' results for satisfiability
in $\StdGroundedVModelOf{\GLinference}$.
In the following, given a term $t$ and set $\cV$ of variables, $\partialGroundingOfWith{t}{\cV}$ is
the term obtained by replacing all variables $x\in\Var(t)\cap\cV$ in $t$ by $c_x$.
Similarly for atoms.
\begin{proposition}
\label{PropUniversalQuantifierEliminationByGrounding}
Let $\GLinference=(\Symbols,\SPredicates,\SetOfInferenceRules)$ be an \ElementaryIS{}
and $A\in\atomsOn{\Symbols,\SPredicates,\Variables}$ be an atom with
variables $x_1,\ldots,x_k\in\Variables$.
Then, 
\vspace{-0.5cm}
\begin{eqnarray}
\satisfactionInThOf{\StdGroundedVModelOf{\GLinference}}{(Q_1 x_1)\cdots(Q_k x_k)\GLatom}
& \text{iff} &
\satisfactionInThOf{\StdGroundedVModelOf{\GLinference}}{(\exists\:x_{\epsilon_1})\cdots(\exists \:x_{\epsilon_p})\partialGroundingOfWith{\GLatom}{\cV_U}}\nonumber
\end{eqnarray}
\vspace{-0.7cm}

\noindent
where, for all $1\leq i\leq k$, $Q_ix_i$ represents a quantified variable $x_i$, where $Q_i$ is a quantifier, either existential ($\exists$) or universal ($\forall$);
$E=\{\epsilon_1,\ldots,\epsilon_p\}$ 
is the set of indices of \emph{existentially quantified} variables;
and $\cV_U$ is the set of universally quantified variables.
\end{proposition}

\begin{theorem}
\label{TheoUniversalQuantifierEliminationPositiveSentencesByGrounding}
Let 
$\GLinference$ be an \ElementaryIS.
Given $n\geq 1$, let $A_1,\ldots,A_n$ be atoms with
variables $x_1,\ldots,x_k$, for some $k\geq 0$.
Given $m\geq 1$ and $1\leq n_i\leq m$ for all $1\leq i\leq m$, let $A_{ij}$ be atoms for all $1\leq i\leq m$ and $1\leq j\leq n_i$ with
variables $x_1,\ldots,x_k$, for some $k\geq 0$. 
Let $Q_q\in\{\exists,\forall\}$ for $1\leq q\leq k$,
$E=\{\epsilon_1,\ldots,\epsilon_p\}=\{q\mid 1\leq q\leq k, Q_q=\exists\}$ 
and $\cV_U$ be the set of universally quantified variables.
Then,
\begin{eqnarray}
\satisfactionInThOf{\StdGroundedVModelOf{\GLinference}}{(\forall\vec{x})\bigwedge_{i=1}^{n}A_{i}}
& \text{ iff }  &
\satisfactionInThOf{\StdGroundedVModelOf{\GLinference}}{\bigwedge_{j=1}^{n}A^\downarrow_{i}}
\label{TheoUniversalQuantifierEliminationECBCAsByGrounding_UQConjunction}
\end{eqnarray}
\begin{eqnarray}
\satisfactionInThOf{\StdGroundedVModelOf{\GLinference}}{(\forall\vec{x})\bigvee_{i=1}^m\bigwedge_{j=1}^{n_i}A_{ij}}
& \text{ if } &
\satisfactionInThOf{\StdGroundedVModelOf{\GLinference}}{\bigvee_{i=1}^m\bigwedge_{j=1}^{n_i}A^\downarrow_{ij}}~\label{TheoUniversalQuantifierEliminationECBCAsByGrounding_UQDNF}
\end{eqnarray}
\begin{eqnarray}
\text{If } \satisfactionInThOf{\StdGroundedVModelOf{\GLinference}}{\overrightarrow{(Q_q x_q)}\bigwedge_{i=1}^{n}A_{i}},
& \text{ then } &
\satisfactionInThOf{\StdGroundedVModelOf{\GLinference}}{(\exists\:x_{\epsilon_1})\cdots(\exists \:x_{\epsilon_p})\bigwedge_{i=1}^{n}\partialGroundingOfWith{A_{i}}{\cV_U}}\label{TheoUniversalQuantifierEliminationECBCAsByGrounding_Conjunction}
\end{eqnarray} 
\begin{eqnarray}
\satisfactionInThOf{\StdGroundedVModelOf{\GLinference}}{\overrightarrow{(Q_q x_q)}\bigvee_{i=1}^m\bigwedge_{j=1}^{n_i}A_{ij}}
& \text{ if } &
\satisfactionInThOf{\StdGroundedVModelOf{\GLinference}}{(\exists\:x_{\epsilon_1})\cdots(\exists \:x_{\epsilon_p})\bigvee_{i=1}^m\bigwedge_{j=1}^{n_i}\partialGroundingOfWith{A_{ij}}{\cV_U}} ~~~~~\label{TheoUniversalQuantifierEliminationECBCAsByGrounding_DNF}
\end{eqnarray}
Finally, if, for all $j\in E$, $x_j$ occurs in at most one $A_i$, for some 
$1\leq i\leq n$, then
\begin{eqnarray}
\satisfactionInThOf{\StdGroundedVModelOf{\GLinference}}{\overrightarrow{(Q_q x_q)}\bigwedge_{i=1}^{n}A_{i}},
& \text{ iff } &
\satisfactionInThOf{\StdGroundedVModelOf{\GLinference}}{(\exists\:x_{\epsilon_1})\cdots(\exists \:x_{\epsilon_p})\bigwedge_{i=1}^{n}\partialGroundingOfWith{A_{i}}{\cV_U}}\label{TheoUniversalQuantifierEliminationECBCAsByGrounding_ConjunctionDisjointExQVars}
\end{eqnarray} 
\end{theorem}

\begin{table}[t]
\caption{Canonical models for Elementary Inference Systems $\GLinference=(\Symbols,\SPredicates,\SetOfInferenceRules)$}
\begin{center}
\begin{tabular}{c@{\hspace{1cm}}c@{\hspace{1cm}}c@{\hspace{1cm}}c@{\hspace{1cm}}c}
Canonical model & Signatures & Type & Atoms in & 
 \\
 \hline
 $\StdModelOf{\GLinference}$ 
 & $\Symbols$, $\SPredicates$ & Herbrand & $\atomsOn{\Symbols,\SPredicates,\emptyset}$
 \\
  $\StdVModelOf{\GLinference}$ 
 & $\Symbols$, $\SPredicates$ & V-Herbrand & $\atomsOn{\Symbols,\SPredicates,\Variables}$
 \\
  $\StdGroundedVModelOf{\GLinference}$ 
 & $\GroundedSymbols$, $\SPredicates$ & Herbrand & $\atomsOn{\GroundedSymbols,\SPredicates,\emptyset}$
 \end{tabular}
\end{center}
\label{TableCanonicalModelsForEIS}
\end{table}

\section{Semantic Properties of Elementary Inference Systems}
\label{SecProvingSemanticPropertiesOfEIS}

In the following, we adapt the definitions in \cite{Lucas_ProvingSemanticPropertiesAsFirstOrderSatisfiability_AI19} 
to the specific setting of \ElementaryIS.

\begin{definition}[Semantic property, cf.\ {\cite[Definition 11]{Lucas_ProvingSemanticPropertiesAsFirstOrderSatisfiability_AI19}}]
\label{DefSemanticPropertyOfAnEIS}
Let $\GLinference=(\Symbols,\SPredicates,\SetOfInferenceRules)$ be an \ElementaryIS{} and $\StdModel{}$ be an $\Symbols',\SPredicates'$-model of $\GLtheoryOf{\GLinference}$ for some $\Symbols'\supseteq\Symbols$ and 
$\SPredicates'\supseteq\SPredicates$ extending $\Symbols$ and $\SPredicates$, respectively.
Then,  $\GLformula\in\formulasOn{\Symbols',\SPredicates',\Variables}$ is a  \emph{semantic property} of $\GLinference$ 
(relative to $\StdModel{}$, or 
just an $\StdModel{}$-property) 
if $\satisfactionInThOf{\StdModel{}}{\GLformula}$.
\end{definition}
\begin{remark}[Use of extended signatures]
\label{RemUseOfExtendedSignatures}
In contrast to \cite[Definition 11]{Lucas_ProvingSemanticPropertiesAsFirstOrderSatisfiability_AI19}, in 
Definition \ref{DefSemanticPropertyOfAnEIS} we consider extensions $\Symbols'$ and $\SPredicates'$ of the original signatures $\Symbols$ and $\SPredicates$ of the considered 
$\ElementaryIS$ because we consider properties expressed as sentences in $\formulasOn{\GroundedSymbols,\SPredicates,\emptyset}$ which must be satisfied in the Herbrand $\GroundedSymbols,\SPredicates$-interpretation $\StdGroundedVModelOf{\GLinference}$, as $\StdModelOf{\GLinference}$ and
$\StdVModelOf{\GLinference}$ provide no interpretation for symbols in $\GroundedSymbols$.
\end{remark}
Many properties of \gtrs{s} $\cR$ can be expressed as \emph{semantic properties} relative to 
$\StdVModelOf{\cR}$ (equivalently $\StdGroundedVModelOf{\cR}$, see Theorem \ref{TheoEquivalenceOfVModelAndGroundedVModel}), 
or $\StdModelOf{\GLinference}$ (for the ground version).
In general, such models are \emph{not} comparable regarding their ability to express properties of \ElementaryIS{s}.
Thus, the appropriate choice of 
a reference model is essential to characterize the targeted property.
The \emph{shape} of formulas $\GLformula$ also plays a role.
We often consider \emph{positive} 
sentences $\GLformula$ 
of the form:
\begin{eqnarray}
(Q_1 x_1)\cdots(Q_k x_k) \bigwedge_{i=1}^m\bigvee_{j=1}^{n_i}  A_{ij}\label{ManySortedClosureOfClauses}
\end{eqnarray} 
where 
(a) for all $1\leq i\leq m$ and $1\leq j\leq n_i$,
$A_{ij}$ are \emph{atoms} (which is the reason why we talk 
of ``\emph{positive}'' formulas), 
(b) $x_1,\ldots,x_k$ for some $k\geq 0$ are the variables occurring in those atoms and
(c) $Q_1,\ldots,Q_k$ are universal/existential quantifiers.
If $Q_i=\exists$ for all $1\leq q\leq k$, we say that (\ref{ManySortedClosureOfClauses}) is an
\emph{Existentially Closed Boolean Combination of Atoms} (\ecbca{} for short).
We have the following.

\begin{proposition}
\label{PropECBCAsAsSemanticProperties}
Let $\GLinference=(\Symbols,\SPredicates,\SetOfInferenceRules)$ be an 
\ElementaryIS{} and $\GLformula\in\formulasOn{\Symbols,\SPredicates,\Variables}$ 
be an \ecbca.
If $\satisfactionInThOf{\StdModelOf{\GLinference}}{\GLformula}$, 
then
$\satisfactionInThOf{\StdVModelOf{\GLinference}}{\GLformula}$
and 
$\satisfactionInThOf{\StdGroundedVModelOf{\GLinference}}{\GLformula}$.
\end{proposition}
Formulas (\ref{ManySortedClosureOfClauses}) where \emph{only conjunction} is used are called \emph{and}- (or $\wedge$-)formulas.

\subsection{Semantic Properties as Logical Consequences}

We can prove semantic properties of \ElementaryIS{} as logical consequences.

\begin{proposition}
\label{PropCorollary14_Lucas19}
(cf.\ \cite[Corollary 14]{Lucas_ProvingSemanticPropertiesAsFirstOrderSatisfiability_AI19})
Let $\GLinference$ be an \ElementaryIS{} and $\StdModel{}$ be a model of $\GLtheoryOf{\GLinference}$.
Every logical consequence 
of $\GLtheoryOf{\GLinference}$ is an
$\StdModel{}$-property of $\GLinference$.
\end{proposition}
In general, this result \emph{cannot} be reversed 
\cite{Lucas_ProvingSemanticPropertiesAsFirstOrderSatisfiability_AI19}.
By Proposition \ref{PropCorollary14_Lucas19}, 
we can use theorem provers,
e.g., \ProverNine{} \cite{McCune_Prove9andMace4_Unpublished10}) to prove semantic properties, 
 although 
without
distinguishing 
different (canonical) models.

\begin{example}
Term
$\fS{s}(\fS{s}(\fS{s}(x)))$ is reducible for arbitrary instances of $x$ to terms in $\Terms$ if
\begin{eqnarray}
\satisfactionInThOf{\StdGroundedVModelOf{\cR}}{(\forall x)(\exists z)\:\fS{s}(\fS{s}(\fS{s}(x)))\rew{}z}\label{LblInstancesOfsssxAreReducible_SemProperty}
\end{eqnarray}
holds.
Since $(\forall x)(\exists z)\:\fS{s}(\fS{s}(\fS{s}(x)))\rew{}z$  is a logical consequence
of $\rewtheoryOf{\cR}$ (use $\ProverNine$), by Proposition \ref{PropCorollary14_Lucas19}, (\ref{LblInstancesOfsssxAreReducible_SemProperty}) holds.
\end{example}
As for Example \ref{ExPEvenZeroOdd},
Proposition \ref{PropCorollary14_Lucas19} cannot be used to prove $\satisfactionInThOf{\rewtheoryOf{\cR}}{(\ref{ExPEvenZeroOdd_TermsEitherPositiveOddZero})}$ for $\rewtheoryOf{\cR}$ in Figure \ref{FigExPEvenZeroOdd_Th}: 
a \emph{model} of $\rewtheoryOf{\cR}\cup\{\neg(\ref{ExPEvenZeroOdd_TermsEitherPositiveOddZero})\}$ can be obtained with, e.g., \MaceFour{} \cite{McCune_Prove9andMace4_Unpublished10}, i.e.,  (\ref{ExPEvenZeroOdd_TermsEitherPositiveOddZero}) is not
a \emph{logical consequence} of
$\rewtheoryOf{\cR}$.

\subsection{Semantic Properties as Inductive Consequences}

For universally quantified positive formulas $\GLformula$
we can prove 
$\satisfactionInThOf{\StdModelOf{\GLinference}}{\GLformula}$ by 
\emph{induction} on the structure of the set of ground terms $\GTerms$.

\begin{example}
\label{ExPEvenZeroOdd_TermsEitherPositiveOddZero_InductiveProof}
For $\cR$ in Example \ref{ExPEvenZeroOdd}, we can prove that 
$\satisfactionInThOf{\StdModelOf{\cR}}{(\ref{ExPEvenZeroOdd_TermsEitherPositiveOddZero})}$ by induction on ground terms $t$ instantiating variable $x$ in (\ref{ExPEvenZeroOdd_TermsEitherPositiveOddZero}):
\begin{itemize}
\item Base case: if $t=\fS{0}$, then $\fS{zero}(t)$ holds by an application of 
$(\RuleHornClause)_{(\ref{ExPEvenZeroOdd_horn5})}$
using reflexivity rule $(\RuleReflexivity)$.
\item Induction: let $t=\fS{s}^{n+1}(\fS{0})$ for some $n\geq 0$ and
let $u=\fS{s}^{n}(\fS{0})$, i.e., $t=\fS{s}(u)$. 
Assume that (the matrix of)
(\ref{ExPEvenZeroOdd_TermsEitherPositiveOddZero}) holds on $u$. 
We consider three cases:
\begin{enumerate}
\item If $\proofInISof{\GLinferenceOf{\cR}}{\fS{zero}(u)}$ holds, then, in order to apply 
$(\RuleHornClause)_{(\ref{ExPEvenZeroOdd_horn5})}$, we need either 
$u=\fS{0}$, so that the reflexivity rule $(\RuleReflexivity)$ permits the use of $(\RuleHornClause)_{(\ref{ExPEvenZeroOdd_horn5})}$,
or else to have $n\div 2>0$ applications of $(\RuleHornClause)_{(\ref{ExPEvenZeroOdd_rule1})}$ to remove all occurrences of $\fS{s}$  
from $u$ to finally obtain $\fS{0}$.
Thus, $n$ must be an even number.
However, the application of $(\RuleHornClause)_{(\ref{ExPEvenZeroOdd_rule1})}$
on $u$ requires that $u=\fS{s}(\fS{s}(u'))$ and that $u'\geq\fS{s}(\fS{0})$, which is possible only if $n$ is an odd number.
We obtain a contradiction.
Thus, it must be $u=\fS{0}$ and $t=\fS{s}(\fS{0})$. We conclude $\proofInISof{\GLinferenceOf{\cR}}{\fS{odd}(t)}$ using $(\RuleHornClause)_{(\ref{ExPEvenZeroOdd_horn4})}$.
\item If $\proofInISof{\GLinferenceOf{\cR}}{\fS{odd}(u)}$ holds, then by reasoning as above, $n$ must be an odd number and hence $n+1$ is a positive even number.
We conclude $\proofInISof{\GLinferenceOf{\cR}}{\fS{peven}(t)}$ using 
$(\RuleHornClause)_{(\ref{ExPEvenZeroOdd_horn3})}$.
\item The case when $\proofInISof{\GLinferenceOf{\cR}}{\fS{peven}(u)}$ holds is handled similarly to conclude $\proofInISof{\GLinferenceOf{\cR}}{\fS{odd}(t)}$.
\end{enumerate}
Thus, (the matrix of)
(\ref{ExPEvenZeroOdd_TermsEitherPositiveOddZero}) holds on $t$, as desired.
\end{itemize}
\end{example}
Inductionless induction methods \cite{Comon_InductionlessInduction_HAR01,ComNiu_InductionIAxiomatizationFirstOrderConsistency_IC00} could also be used, as they provide a way to reduce proofs of inductive consequence \cite[Definition 2.1]{Comon_InductionlessInduction_HAR01} 
(which implies satisfiability in the least Herbrand model) to proofs of consistency. 
A set $\Iaxiomatization$ of first-order formulas is an $I$-axiomatization of 
the minimal model $\StdModelOf{\GLtheory}$ of a Horn theory $\GLtheory$ if
(i) $\Iaxiomatization$ is a recursive set and contains only purely universal sentences and
(ii) $\StdModelOf{\GLtheory}$ is the only Herbrand model of $\GLtheory\cup\Iaxiomatization$ up to isomorphism
\cite[Definition 3]{ComNiu_InductionIAxiomatizationFirstOrderConsistency_IC00}.
Then, we have:

\begin{proposition}[{\cite[Proposition 7]{ComNiu_InductionIAxiomatizationFirstOrderConsistency_IC00}}]
\label{Prop_ComNiu00_Proposition7}
Let $\Iaxiomatization$ be an $I$-axiomatization of $\StdModelOf{\GLtheory}$
and $C$ be a  set of clauses.
Then, $\Iaxiomatization\cup\GLtheory\cup C$ is H-consistent iff $\satisfactionInThOf{\StdModelOf{\GLtheory}}{C}$.
\end{proposition}
In general, Proposition \ref{Prop_ComNiu00_Proposition7} cannot be used with 
existentially quantified sentences $\GLformula$ as the standard clausal form $C_\GLformula$
would require skolemization which neither preserve H-consistency 
(see Example \ref{ExSkolemizationAndHinconsistency}) 
nor satisfiability in a given structure (in this case $\StdModelOf{\GLtheory}$),
see Example \ref{ExSkolemizationAndLogicalEquivalence}.
By \cite[Theorem 4.2]{ChaLee_symbolicLogicAndMechanicalTheoremProving_1973}, consistency and H-consistency are equivalent for clauses.
Thus, we have:

\begin{corollary}
\label{Coro_ComNiu00_Proposition7}
Let $\Iaxiomatization$ be an $I$-axiomatization of $\StdModelOf{\GLtheory}$
and $C$ be a set of clauses.
Then, $\Iaxiomatization\cup\GLtheory\cup C$ is consistent iff  $\satisfactionInThOf{\StdModelOf{\GLtheory}}{C}$.
\end{corollary}
However, obtaining appropriate $I$-axiomatizations can be difficult.

\subsection{Using Satisfiability in Arbitrary Interpretations}
\label{SecUsingSatisfiabilityInArbitraryInterpretations}
Satisfiability in a canonical model can be undecidable (as the membership relation is based on provability or deduction).
As in
\cite{Lucas_ProvingSemanticPropertiesAsFirstOrderSatisfiability_AI19},
we show how to use satisfaction in arbitrary first-order interpretations $\SStructure$.
Given
$\Symbols,\SPredicates$-structures $\SStructure$ and $\SStructure'$, a mapping $h:\domainOf{\SStructure}\to\domainOf{\SStructure'}$ (or just $h:\SStructure\to\SStructure'$ if no confusion arises) is a \emph{homomorphism} if 
(i) for all $k$-ary symbols $f\in\Symbols$ and all $a_1,\ldots,a_k\in\SStructure$, $h(f^\SStructure(a_1,\ldots,a_k))=f^{\SStructure'}(h(a_1),\ldots,h(a_k))$
and 
(ii) for all $n$-ary predicates $P\in\SPredicates$ and $a_1,\ldots,a_n\in\SStructure$, if $P^\SStructure(a_1,\ldots,a_n)$ holds,
then $P^{\SStructure'}(h(a_1),\ldots,h(a_n))$ holds as well
\cite[Theorem 1.3.1(a) \& (b)]{Hodges_AShorterModelTheory_1997}.
Every model $\SStructure$ of a 
set $\HerbrandInterpretation\subseteq\atomsOn{\Symbols,\SPredicates,\emptyset}$ of \emph{ground atoms} 
has a \emph{unique} homomorphism
$h:\GTerms\to\SStructure$ 
\cite[Theorem 1.5.1]{Hodges_AShorterModelTheory_1997}
(the so-called \emph{interpretation homomorphism}).
Remind that a mapping $f:D\to E$ is \emph{surjective} if for all $y\in E$ there is $x\in D$ such that $f(x)=y$.

\begin{theorem}[Disproving positive $\StdModelOf{\GLinference}$-properties] 
\label{TheoDisprovingSemanticPropertiesWrtHerbrandModel}
(cf. \cite[Corollary 28]{Lucas_ProvingSemanticPropertiesAsFirstOrderSatisfiability_AI19})
Let 
 $\GLinference=(\Symbols,\SPredicates,\SetOfInferenceRules)$ be an \ElementaryIS, 
 $\GLformula\in\formulasOn{\Symbols,\SPredicates,\Variables}$ be a 
 positive 
 sentence (\ref{ManySortedClosureOfClauses}), and  
 $\SStructure$ be an $\Symbols,\SPredicates$-structure
 satisfying $\GLtheoryOf{\GLinference}\cup\{\neg\GLformula\}$.
If (i) $\GLformula$ is an \ecbca,
or (ii) $h:\GTerms\to\SStructure$ is surjective, then 
$\satisfactionInThOf{\StdModelOf{\GLinference}}{\neg\GLformula}$ holds.
\end{theorem}
Models $\SStructure$ required
in Theorem
\ref{TheoDisprovingSemanticPropertiesWrtHerbrandModel} can often be automatically 
generated 
by using
model generators like \AGES{} 
\cite{GutLuc_AutomaticGenerationOfLogicalModelsWithAGES_CADE19}
or \MaceFour{}
\cite{McCune_Prove9andMace4_Unpublished10}.

\begin{example}
The following \ecbca{} represents the existence of a \emph{cycle} in
rewriting computations:
\begin{eqnarray}
(\exists x)(\exists y)~x\rew{}y\wedge y\rews{}x\label{LblRnoCycling}
\end{eqnarray}
We prove that no \emph{ground} term starts a cycling reduction with $\cR$ in Example \ref{ExPEvenZeroOdd}.
By Theorem \ref{TheoDisprovingSemanticPropertiesWrtHerbrandModel}.(i), 
we need to show that
there is a model $\SStructure$ of $\rewtheoryOf{\cR}$ which also satisfies $\neg(\ref{LblRnoCycling})$.
We use \AGES{} to find such a model: the domain is $\SStructure=\{z\in\integers\mid z\leq 1\}$; function and predicate symbols are interpreted as follows:
 \[\begin{array}{rcl@{\hspace{1cm}}rcl@{\hspace{1cm}}rcl@{\hspace{1cm}}rcl}
\fS{0}^\SStructure & = & 1
&
\fS{s}^\SStructure(x) & = & x-1
\\
\fS{odd}^\SStructure(x) & \Leftrightarrow & true
&
\fS{peven}^\SStructure(x) & \Leftrightarrow & true
&
\fS{zero}^\SStructure(x) & \Leftrightarrow & true
\\
x\geq^\SStructure y & \Leftrightarrow & true
&
x\rew{}^\SStructure y & \Leftrightarrow & y>x
&
x(\rews{})^\SStructure y & \Leftrightarrow & y\geq x
\end{array}
\]
\end{example}
Surjectivity of $h:\GTerms\to\SStructure$ (required in Theorem \ref{TheoDisprovingSemanticPropertiesWrtHerbrandModel}(ii))
can be guaranteed by using 
an appropriate theory  
$\SurHomTh$
\cite[Section 6]{Lucas_ProvingSemanticPropertiesAsFirstOrderSatisfiability_AI19}.
For instance, given a non-empty, finite set $T\subseteq\GTerms$ of ground terms and
\begin{eqnarray*}
\SurHomTh^T=\{(\forall x)\bigvee_{t\in T}x=t\}
\end{eqnarray*}
by \cite[Proposition 40]{Lucas_ProvingSemanticPropertiesAsFirstOrderSatisfiability_AI19},
$\satisfactionInThOf{\SStructure}{\SurHomTh^T}$
implies surjectivity of $h$.
A more general approach is described in \cite[Section 6.2]{Lucas_ProvingSemanticPropertiesAsFirstOrderSatisfiability_AI19}.

Formulas $\GLformula$ involving symbols in $\GroundedVariables$ 
cannot be proved as 
semantic properties w.r.t.\ $\StdModelOf{\GLinference}$ because
 symbols in $\GroundedVariables$ are not interpreted by $\StdModelOf{\GLinference}$.
Instead, $\StdGroundedVModelOf{\GLinference}$ should be used.
However, $\StdGroundedVModelOf{\GLinference}$ is an 
$\GroundedSymbols,\SPredicates$-structure.
Hence, $\SStructure$ should be an $\GroundedSymbols,\SPredicates$-structure 
to be able to use Theorem \ref{TheoDisprovingSemanticPropertiesWrtHerbrandModel} applied to $\GroundedSymbols$.
However, $\GroundedSymbols$ is infinite (due to infiniteness of $\Variables$), 
and
synthesizing structures $\SStructure$ interpreting infinitely many symbols
can be difficult.
Since $\GLformula$ contains a \emph{finite} (possibly empty) set of symbols $K\subseteq\GroundedVariables$, and $\GLtheoryOf{\GLinference}\subseteq\formulasOn{\Symbols,\SPredicates,\Variables}$, we can try to use $\Symbols\cup K,\SPredicates$-structures $\SStructure$ instead.

\begin{theorem}[Disproving positive $\StdGroundedVModelOf{\GLinference}$-properties]
\label{TheoDisprovingSemanticPropertiesWrtVGroundedHerbrandModel}
Let $\GLinference=(\Symbols,\SPredicates,\SetOfInferenceRules)$ be an \ElementaryIS,
$\Variables$ be a set of variables,
$K\subseteq\GroundedSymbols$,
and 
 $\GLformula\in\formulasOn{\Symbols\cup K,\SPredicates,\Variables}$ be a positive sentence (\ref{ManySortedClosureOfClauses}), and 
  $\SStructure$ be an $\Symbols\cup K,\SPredicates$-model of $\GLtheoryOf{\GLinference}$.
If 
(i) 
$\GLformula$ is an \ecbca{} and $\satisfactionInThOf{\SStructure}{\neg\GLformula}$ holds,
or 
(ii) 
$\GLformula$ is an $\wedge$-positive formula and $U$ is the set of universally quantified variables in $\GLformula$ and $\satisfactionInThOf{\SStructure}{\neg\partialGroundingOfWith{\GLformula}{U}}$ holds
or 
(iii) $h:\GTermsOn{\Symbols\cup K}\to\SStructure$ is surjective and 
$\satisfactionInThOf{\SStructure}{\neg\GLformula}$ holds, 
then $\satisfactionInThOf{\StdGroundedVModelOf{\GLinference}}{\neg\GLformula}$ holds.
\end{theorem}
\begin{remark}[Formulas 
$\GLformula\in\formulasOn{\Symbols,\SPredicates,\Variables}$ without grounded variables]
\label{RemProvingFormulasWithoutGroundedVariablesAsSemanticPropertiesOfGroundedModel}
If $\GLformula$ 
contains no grounded variables $c_x$, then $K$ in Theorem 
\ref{TheoDisprovingSemanticPropertiesWrtVGroundedHerbrandModel} can be taken as 
\emph{empty}.
In this case, proving that $\satisfactionInThOf{\StdGroundedVModelOf{\GLinference}}{\neg\GLformula}$ holds using items (i) and (iii) in Theorem \ref{TheoDisprovingSemanticPropertiesWrtVGroundedHerbrandModel}
would also prove $\satisfactionInThOf{\StdModelOf{\GLinference}}{\neg\GLformula}$ as those
items would 
coincide with the conditions of use of Theorem \ref{TheoDisprovingSemanticPropertiesWrtHerbrandModel}.
However, it may happen that 
$\satisfactionInThOf{\StdModelOf{\GLinference}}{\GLformula}$ holds
but 
$\satisfactionInThOf{\StdGroundedVModelOf{\GLinference}}{\GLformula}$ does \emph{not} hold
(see Example \ref{ExPEvenZeroOdd_DifferencesInCanonicalModels}
and Example \ref{ExPEvenZeroOdd_TermsEitherPositiveOddZero_NonGroundDisproved} below).
In this case, with $K=\emptyset$, Theorem \ref{TheoDisprovingSemanticPropertiesWrtVGroundedHerbrandModel} could \emph{not} be used to conclude $\satisfactionInThOf{\StdGroundedVModelOf{\GLinference}}{\neg\GLformula}$.
Then, we let $K\neq\emptyset$ so that Theorem  \ref{TheoDisprovingSemanticPropertiesWrtVGroundedHerbrandModel} can be 
advantageously used.
\end{remark}
\begin{example}
\label{ExPEvenZeroOdd_TermsEitherPositiveOddZero_NonGroundDisproved}
We prove that $\satisfactionInThOf{\StdGroundedVModelOf{\cR}}{\neg(\ref{ExPEvenZeroOdd_TermsEitherPositiveOddZero})}$ holds by using Theorem \ref{TheoDisprovingSemanticPropertiesWrtVGroundedHerbrandModel}.(iii).
Let $K=\{c_x\}$ 
and $T=\{\fS{0},c_x\}$. 
Hence, $\SurHomTh^T=\{(\forall x)\:x=\fS{0}\vee x=c_x\}$.
We obtain a
model $\SStructure$ of
\[\rewtheoryOf{\cR}\cup\SurHomTh^T\cup\{\neg(\forall x)(\fS{peven}(x)\vee\fS{odd}(x))\vee\fS{zero}(x))\}
\] 
with \MaceFour.
The domain 
is $\{0,1\}$; 
the interpretations of function
symbols is 
\[
\begin{array}{rcl@{\hspace{2cm}}rcl@{\hspace{2cm}}rcl@{\hspace{1cm}}rcl}
\fS{c_x}^\SStructure  & = &  1
&
\fS{0}^\SStructure & = & 1
&
\fS{s}^\SStructure(x) & = &  x+1
\end{array}
\]
and all predicate symbols (except the equality symbol) 
are interpreted as \emph{true}.
\end{example}

\section{Related work}
\label{SecRelatedWork}

Our elementary inference systems combine aspects of
Smullyan's \emph{Elementary Formal Systems} 
and Mathematical Systems \cite[Chapter 1, \#A, \textsection 1 and \textsection 4]{Smullyan_TheoryOfFormalSystems_1961} 
(emphasizing the idea of \emph{defining} sets or relations by deduction using implicative (schemes of) axioms $B_1\Rightarrow B_2\Rightarrow\cdots\Rightarrow B_n\Rightarrow B$, where $B_1,\ldots,B_n$ and $B$ are \emph{atoms})
and 
Gentzen's notion of \emph{inference rules} (where such implicative axioms are displayed as inference rules $\frac{B_1 \cdots  B_n}{B}$) and the 
arrangement of deductions as \emph{formula-trees}
\cite[Chapter 1, \textsection 2, B]{Prawitz_NaturalDeductionAProofTheoreticalStudy_1965}, which is essential to make sense of the notion of \emph{operational (non-)termination}, which cannot be captured by using Smullyan's notion of deduction of atoms in an elementary formal system.
On the other hand, Gentzen's general notion of inference rule 
$\frac{\GLformula_1\cdots\GLformula_n}{\GLformula}$ (or \emph{inference figure} in his terminology) permits the use of arbitrary formulas 
$\GLformula_1,\ldots,\GLformula_n$ and $\GLformula$ in the upper and lower parts of the inference rule \cite[Section I, item 3.1]{Gentzen_UntersuchungenUberDasLogischeSchliessenPart1_MZ35}, thus obtaining
\emph{more general} inference rules 
than Smullyan's and ours. 
Both Smullyan and Prawitz emphasize the use of \emph{instances} of inference rules in deduction (rather than the explicit inclusion of substitutions in rules, as in \cite{BruMes_SemanticFoundationsForGeneralizedRewriteTheories_TCS06,Lalement_ComputationAsLogic_1993}) a keypoint which we follow in our definitions and methods.

After describing a computational system as a \emph{first-order theory} $\GLtheory$, 
the use of first-order sentences $\GLformula$ 
to express properties of a computational system (programming language, database, etc.) is a natural choice \cite{GreRap_TheUseOfTheoremProvingTechniquesInQuestionAnsweringSystems_ACMNC68,Manna_PropertiesOfProgramsAndTheFirstOrderLogicCalculus_JACM69},
and a \emph{``properties-as-logical-consequences''} 
approach has been frequently adopted to claim/deny the property of the considered system \cite{GreRap_TheUseOfTheoremProvingTechniquesInQuestionAnsweringSystems_ACMNC68}.
Clark's approach, however, is that
sentences expressing 
properties
should be checked \emph{with respect to a given canonical model
only} 
\cite[Chapter 4]{Clark_PredicateLogicAsAComputationalFormalism_TR79}.
After the seminal work on the model-theoretic description of the semantics of 
logic programming \cite{EmdKow_TheSemanticsOfPredicateLogicAsAProgrammingLanguage_JACM76},
other approaches have been proposed, including the use of non-ground Herbrand interpretations \cite{Clark_PredicateLogicAsAComputationalFormalism_TR79}
and other refinements \cite{FalLevMarPal_AModelTheoreticReconstructionOfTheOperationalSemanticsOfLogicPrograms_IC93,%
FalLevPalMar_DeclarativeModelingOfTheOperationalBehaviorOfLogicLanguages_TCS89,%
LevPal_TheDeclarativeSemanticsOfLogicalReadOnlyVariables_SLP85}.
In the realm of Term Rewriting Systems, 
a different path has been followed using the
\emph{first-order theory of rewriting} (\emph{FOThR}) 
for \trs{s} $\cR$ 
\cite{DauTis_TheTheoryOfGroundRewriteSystemsIsDecidable_LICS90}, where
predicate symbols $\to$ and $\to^*$ are \emph{interpreted} on 
the least Herbrand model $\StdModelOf{\cR}$ of $\rewtheoryOf{\cR}$.
However, only
formulas $\GLformula$ containing \emph{no constant or function symbol} 
can be used to express properties which are checked by satisfiability in
$\StdModelOf{\cR}$ \cite[Section 6]{Dauchet_RewritingAndTreeAutomata_TCSschool93}.
For instance, \emph{ground confluence} of rewriting computations is expressed as follows:
\begin{eqnarray}
(\forall x,y,z)~(x\to^*y\wedge x\to^*z\Rightarrow (\exists u) (y \to^*u\wedge z\to^*u))\label{FOThRsentenceForConfluence}
\end{eqnarray}
and
$\satisfactionInThOf{\StdModelOf{\cR}}{(\ref{FOThRsentenceForConfluence})}$ means that
$\cR$ is ground confluent, 
as variables in (\ref{FOThRsentenceForConfluence}) range on \emph{ground} terms
(the Herbrand Universe) only.
Tree automata techniques can be used to prove properties of \emph{ground} TRSs
$\cR$.
Recently,  the approach was extended to left-linear, right-ground TRSs
\cite{RapMid_AutomatingTheFirstOrderTheoryOfRewritingForLeftLinearRightGroundRewriteSystems_FSCD16}.
The tool \FORT{} \cite{RapMid_FORT2_0_IJCAR18} provides an implementation.
In contrast, 
we are able to deal with \gtrs{s} and properties can be expressed in a more flexible way.
For instance, among the properties
considered above, 
only non-cyclingness of $\cR$ (\ref{LblRnoCycling}) 
can be expressed in FOThR;
however, the results in \cite{DauTis_TheTheoryOfGroundRewriteSystemsIsDecidable_LICS90,RapMid_AutomatingTheFirstOrderTheoryOfRewritingForLeftLinearRightGroundRewriteSystems_FSCD16} does not apply to prove it of 
$\cR$ in Example \ref{ExPEvenZeroOdd}.

\section{Conclusions and Future Work}
\label{SecConclusions}

Borrowing Smullyan's \emph{elementary formal systems} using Gentzen's notation for inference rules we have introduced \emph{Elementary Inference Systems} (\ElementaryIS{s}) $\GLinference$, consisting of (elementary) inference rules $\frac{B_1,\ldots,B_n}{B}$ where  $B$, $B_1,\ldots,B_n$ are atoms.
Sets, relations, and computations can be defined by associating a proof-tree to a given
atom $A=P(t_1,\ldots,t_n)$ which is \emph{matched} by the lower part $B$ of an inference rule $\frac{B_1,\ldots,B_n}{B}$, i.e., $A=\sigma(B)$ for some substitution $\sigma$, 
provided that the corresponding instances $\sigma(B_i)$ of each $B_i$, $1\leq i\leq n$ can also be proved analogously.
A first-order (Horn) theory $\GLtheoryOf{\GLinference}$ is given to $\GLinference$ so that
atoms $A$ that can be proved in $\GLinference$ can be \emph{deduced} from $\GLtheoryOf{\GLinference}$ and vice versa.
Also, canonical (Herbrand or V-Herbrand) models $\StdModelOf{\GLinference}$, 
$\StdVModelOf{\GLinference}$, 
and $\StdGroundedVModelOf{\GLinference}$ of $\GLtheoryOf{\GLinference}$
are given to $\GLinference$ so that properties of $\GLinference$ expressed as first-order sentences $\GLformula$ can often be proved of $\GLinference$ by \emph{satisfaction} in the
corresponding canonical models.
We call them \emph{semantic properties} of $\GLinference$.
Practical and mechanizable approaches to prove semantic properties, including the use of theorem provers and model generation tools like \AGES, \MaceFour, and \ProverNine, have been 
illustrated by means of examples showing their use in the analysis of semantic properties of \gtrs{s}.
In the future, we intend to give direct support in \AGES{} to the techniques described in this paper.

\paragraph{Acknoledgements.} I thank the anonymous reviewers for their useful comments and suggestions.

\bibliographystyle{eptcs}
\bibliography{myBibliography}
\end{document}

%% file: packages.tex
\usepackage{hyperref}

\usepackage{ucs}
\usepackage{inputenc}
\usepackage[ruled,lined]{algorithm2e}

\usepackage{listings}
\usepackage[colorinlistoftodos,prependcaption]{todonotes}

\usepackage{textcomp}
\usepackage{cancel}

\usepackage{tikz}
\usetikzlibrary{patterns}

\usepackage{graphicx}

\usepackage{amsmath,amssymb,latexsym}

\usepackage{etex}
\usetikzlibrary{calc} 						
\usetikzlibrary{shapes,arrows}				
\usetikzlibrary{shapes.multipart}			
\usetikzlibrary{positioning}
\usepackage[all,cmtip]{xy}					
\usepackage{tikz-qtree,tikz-qtree-compat}	
\usepackage{mathtools}
\usepackage{url} 
\usepackage{graphics}

\usepackage{wrapfig}
\usepackage{accents} 

%% file: macros.tex
\def\defemb#1#2{\expandafter\def\csname #1\endcsname
{\relax\ifmmode #2\else\hbox{$#2$}\fi}}
\defemb{cA}{\mathcal{A}}
\defemb{cB}{\mathcal{B}}
\defemb{cC}{\mathcal{C}}
\defemb{cD}{\mathcal{D}}
\defemb{cE}{\mathcal{E}}
\defemb{cF}{\mathcal{F}}
\defemb{cG}{\mathcal{G}}
\defemb{cH}{\mathcal{H}}
\defemb{cI}{\mathcal{I}}
\defemb{cJ}{\mathcal{J}}
\defemb{cK}{\mathcal{K}}
\defemb{cL}{\mathcal{L}}
\defemb{cM}{\mathcal{M}}
\defemb{cN}{\mathcal{N}}
\defemb{cO}{\mathcal{O}}
\defemb{cP}{\mathcal{P}}
\defemb{cQ}{\mathcal{Q}}
\defemb{cR}{\mathcal{R}}
\defemb{cS}{\mathcal{S}}
\defemb{cT}{\mathcal{T}}
\defemb{cU}{\mathcal{U}}
\defemb{cV}{\mathcal{V}}
\defemb{cX}{\mathcal{X}}
\defemb{cZ}{\mathcal{Z}}

\newtheorem{theorem}{Theorem}
\newtheorem{proposition}[theorem]{Proposition}
\newtheorem{corollary}[theorem]{Corollary}
\newtheorem{definition}[theorem]{Definition}
\newtheorem{example}[theorem]{Example}
\newtheorem{remark}[theorem]{Remark}

\newcommand{\naturals}{\mathbb{N}}
\newcommand{\integers}{\mathbb{Z}}

\newcommand{\domainOf}[1]{{\text{dom}(#1)}}

\newcommand{\SStructure}{\cA}

\newcommand{\StdModel}[1]{\cM_{#1}} 

\newcommand{\SemDomain}{\cA}

\newcommand{\HerbrandBaseOf}[1]{\cB_{#1}} 
\newcommand{\HerbrandInterpretation}{\cH} 
\newcommand{\VHerbrandInterpretation}{\widehat{\cH}} 

\newcommand{\HerbrandBase}{\cB} 
\newcommand{\VHerbrandBase}{\widehat{\cB}} 
\newcommand{\VHerbrandBaseOf}[1]{\widehat{\cB}_{#1}} 
\newcommand{\StdModelOf}[1]{{\cM(#1)}} 
\newcommand{\StdVModelOf}[1]{{\widehat{\cM}(#1)}} 
\newcommand{\StdGroundedVModelOf}[1]{{\cM^\downarrow(#1)}}

\newcommand{\AGES}{\mbox{\sf AGES}}
\newcommand{\MaceFour}{\mbox{\sf Mace4}}

\newcommand{\ProverNine}{\mbox{\sf Prover9}}

\newcommand{\TTTTwo}{\textsf{T\kern-0.15em\raisebox{-0.55ex}T\kern-0.15emT\kern-0.15em\raisebox{-0.55ex}2}}

\newcommand{\FORT}{\textsf{Fort}}

\newcommand{\Var}{{\cV}ar} 

\newcommand{\bigfrac}[2]{
\begin{array}[b]{c}
\displaystyle #1\\\hline\displaystyle #2
\end{array}}
\newcommand{\bigfracLabel}[2]{
\begin{array}{c}
\displaystyle #1\\\hline\displaystyle #2
\end{array}}
\newcommand{\bigfracn}[3]{
\begin{array}[b]{c}
\displaystyle #1 \\\hline\displaystyle #2
\end{array}
\hbox to 0pt{\raisebox{0.7em}{{\tiny (#3)}}}
}

\newcommand{\ol}[1]{\overline{#1}}  

\newcommand{\rootTerm}{\mathit{root}}

\newcommand{\Symbols}{{\cF}}

\newcommand{\Variables}{{\cX}}

\newcommand{\TermsOn}[2]{{\cT(#1,#2)}}
\newcommand{\GTermsOn}[1]{{\cT(#1)}}

\newcommand{\Terms}{{\TermsOn{\Symbols}{\Variables}}}

\newcommand{\GTerms}{{\cT(\Symbols)}}

\newcommand{\restrictTo}[2]{#1\!\!\downharpoonright_{#2}}

\newcommand{\GroundedVariables}{{C_\Variables}}
\newcommand{\GroundedSymbols}{{\Symbols_\Variables}}
\newcommand{\GroundedTerms}{\GTermsOn{\GroundedSymbols}}

\newcommand{\grounding}[1]{{{#1}^\downarrow}}

\newcommand{\ungrounding}[1]{{{#1}^\uparrow}}

\newcommand{\partialGroundingOfWith}[2]{#1^{\downarrow_{#2}}}

\newcommand{\sos}{\text{SOS}}

\newcommand{\trs}{\text{TRS}} 

\newcommand{\ctrs}{\text{CTRS}}

\newcommand{\gtrs}{\text{GTRS}}

\newcommand{\Rmaps}[1]{{M_{#1}}}

\newcommand{\muBot}{{\mu_\bot}}

\newcommand{\SurHomTh}{\mathsf{SuH}}

\newcommand{\SPredicates}{{\Pi}}

\newcommand{\sentenceFrom}[1]{\ol{#1}}

\newcommand{\RuleHornClause}{\text{HC}}

\newcommand{\RuleReflexivity}{\text{Rf}}
\newcommand{\RuleCompatibility}{\text{Co}} 
\newcommand{\RulePropagation}{\text{Pr}} 

\newcommand{\GLatom}{A}

\newcommand{\GLformula}{F}
\newcommand{\GLgoal}{G}

\newcommand{\GLinference}{\cI}
\newcommand{\GLinferenceOf}[1]{\GLinference(#1)}

\newcommand{\atomsOn}[1]{\mathit{Atoms}_{#1}}
\newcommand{\formulasOn}[1]{\mathit{Forms}_{#1}}

\newcommand{\ecbca}{\text{ECBCA}}

\newcommand{\ElementaryIS}{\text{EIS}}

\newcommand{\SetOfInferenceRules}{I}

\newcommand{\proofInISof}[2]{{\vdash_{#1}#2}}

\newcommand{\GLtheory}{\mathsf{Th}}
\newcommand{\GLtheoryOf}[1]{{\GLtheory(#1)}}

\newcommand{\rewtheoryOf}[1]{{\ol{#1}}}

\newcommand{\Iaxiomatization}{\mathsf{A}}

\newcommand{\deductionInThOf}[2]{{#1\vdash #2}}
\newcommand{\satisfactionInThOf}[2]{{#1\models #2}}
\newcommand{\satisfactionInInterpretationOf}[2]{{#1\models #2}}

\newcommand{\rootTree}{\mathit{root}}

\newcommand{\lhsr}{\ell}
\newcommand{\rhsr}{r}
\newcommand{\gencond}{c}

\newcommand{\IF}{\Leftarrow}
\newcommand{\IFhorn}{\Leftarrow}

\newcommand{\genrelation}{\mathsf{R}}

\newcommand{\rew}[1]{\to_{#1}}

\newcommand{\rewp}[1]{\to^+_{#1}}
\newcommand{\rews}[1]{\to^*_{#1}}

\newcommand{\leftarrown}[1]{{{\:}^n\!\!\leftarrow}}

\newcommand{\leftarrowp}[1]{{{\:}^+\!\!\leftarrow}}
\newcommand{\leftarrows}[1]{{{\:}^*\!\!\leftarrow}}

\pgfdeclarepatternformonly{VGrid}
{
  \pgfpointorigin
}
{
  \pgfqpoint{0.20cm}{0.20cm}
}
{  \pgfqpoint{0.20cm}{0.20cm}
}
{
  \pgfsetlinewidth{0.3pt}
  \pgfpathmoveto{\pgfqpoint{0cm}{0cm}}
  \pgfpathlineto{\pgfqpoint{0cm}{0.20cm}}
  \pgfusepath{stroke}
}

\pgfdeclarepatternformonly{HVGrid}
{
  \pgfpointorigin
}
{
  \pgfqpoint{0.20cm}{0.20cm}
}
{  \pgfqpoint{0.20cm}{0.20cm}
}
{
  \pgfsetlinewidth{0.3pt}
  \pgfpathmoveto{\pgfqpoint{0cm}{0cm}}
  \pgfpathlineto{\pgfqpoint{0cm}{0.20cm}}
  \pgfpathmoveto{\pgfqpoint{0cm}{0cm}}
  \pgfpathlineto{\pgfqpoint{0.20cm}{0cm}}
  \pgfusepath{stroke}
}

\newcommand{\ignore}[1]{}
\newcommand{\ignoreproofs}[1]{}

\newcommand{\fS}[1]{\mathsf{#1}}

\tikzstyle{decision} = [diamond, draw, fill=yellow!20, text width=5em, text badly centered, minimum height=4em, inner sep=0pt, aspect=2]
\tikzstyle{block} = [rectangle, draw,fill=blue!20, text width=5em, text centered, minimum height=4em, rounded corners]
\tikzstyle{cloud} = [ellipse, draw,fill=red!20, text width=5em, text centered, minimum height=4em]
\tikzstyle{line} = [draw, -latex']
\tikzstyle{blockR} = [rectangle, draw, fill=red!20, text centered, minimum height=4em, rounded corners, minimum height=0.75cm]
\tikzstyle{blockB} = [rectangle, draw, fill=blue!20, text centered, minimum height=4em, rounded corners, minimum height=0.75cm]
\tikzstyle{blockG} = [rectangle, draw, fill=green!20, text centered, minimum height=4em, rounded corners, minimum height=0.75cm]
\tikzstyle{blockY} = [rectangle, draw, fill=yellow!20, text centered, minimum height=4em, rounded corners, minimum height=0.75cm]
\tikzstyle{blockW} = [rectangle, draw, fill=white!20, text centered, minimum height=4em, rounded corners, minimum height=0.75cm]
\tikzstyle{blockK} = [rectangle, draw, fill=gray!20, text centered, minimum height=4em, rounded corners, minimum height=0.75cm]
\tikzstyle{mblockB} = [rectangle, draw, fill=blue!20, text centered, minimum height=4em, double,rounded corners, minimum height=0.75cm]
\tikzstyle{mblockW} = [rectangle, draw, fill=white!20, text centered, minimum height=4em, double,rounded corners, minimum height=0.75cm]
\tikzstyle{mblockR} = [rectangle, draw, fill=red!20, text centered, minimum height=4em, double,rounded corners, minimum height=0.75cm]
\tikzstyle{mblockG} = [rectangle, draw, fill=green!20, text centered, minimum height=4em, double,rounded corners, minimum height=0.75cm]

\tikzstyle{circleY} = [circle, draw, fill=yellow!20, text centered, minimum height=4em, rounded corners, minimum height=0.75cm]
\tikzstyle{circleR} = [circle, draw, fill=red!20, text centered, minimum height=4em, rounded corners, minimum height=0.75cm]
\tikzstyle{circleG} = [circle, draw, fill=green!20, text centered, minimum height=4em, rounded corners, minimum height=0.75cm]
\tikzstyle{circleW} = [circle, draw, fill=white!20, text centered, minimum height=4em, rounded corners, minimum height=0.75cm]
\tikzstyle{triangleW} = [isosceles triangle, draw, fill=white!20, text centered, shape border rotate = 90, isosceles triangle stretches]
\tikzstyle{triangleG} = [isosceles triangle, draw, fill=green!20, text centered, shape border rotate = 90, isosceles triangle stretches]
\tikzstyle{triangleR} = [isosceles triangle, draw, fill=red!20, text centered, shape border rotate = 90, isosceles triangle stretches]